\documentclass[aps,prx,floatfix,nopacs,superscriptaddress,twocolumn,reprint]{revtex4-2}

\usepackage[utf8]{inputenc}
\usepackage[american,]{babel}
\usepackage[T1]{fontenc}
\usepackage[pdftex]{graphicx}  
\usepackage{graphicx, xcolor}
\usepackage{amsmath,amsthm,amssymb}
\usepackage{hyperref}
\usepackage{xcolor}
\usepackage{subfigure}
\hypersetup{colorlinks,bookmarksopen,bookmarksnumbered,
citecolor=[rgb]{0.255,0.412,0.882},
linkcolor=[rgb]{0.255,0.412,0.882},
pdfstartview=false,
urlcolor=[rgb]{0.255,0.412,0.882}}
\usepackage{soul}

\begin{document}

\title{Quenches and (Pre)Thermalisation in a mixed Sachdev-Ye-Kitaev Model}

\author{Ancel Larzul}
\author{Marco Schir\'o}\thanks{ On Leave from: Institut de Physique Th\'{e}orique, Universit\'{e} Paris Saclay, CNRS, CEA, F-91191 Gif-sur-Yvette, France}
\affiliation{JEIP, USR 3573 CNRS, Coll\`{e}ge de France,   PSL  Research  University, 11,  place  Marcelin  Berthelot,75231 Paris Cedex 05, France}

\begin{abstract}
We study the nonequilibrium quench dynamics of a mixed Sachdev-Ye-Kitaev model, with competing two bodies random interactions leading to maximally chaotic Non-Fermi Liquid dynamics  and a single body term which dominates at low temperatures and leads to Fermi liquid behavior. For different quench protocols, including sudden switching of two-body interaction and double quench protocols, we solve the large $N$ real-time Dyson equation on the Keldysh contour and compute the dynamics of Green's functions from which we obtain effective temperature and relaxation rates. We show that the model thermalizes to a finite temperature equilibrium and that depending on the value of the quench parameters the effective temperature can be below or above the Fermi-Liquid to Non-Fermi Liquid crossover scale, which can then be accessed through the nonequilibrium dynamics. We identify quench protocols for which the heating dynamics slow down significantly, an effect that we interpret as a signature of prethermalization.
\end{abstract}

\date{\today}
\maketitle

\section{Introduction}

Generic interacting quantum many body systems are expected to thermalize when let evolved unitarily under the action of their own Hamiltonian.  The understanding of this quantum thermalization process, its possible slowdown or complete breakdown is still a subject of large interest and effort, in a broad community ranging from high-energy physics to condensed matter, atomic physics and quantum information.  A particularly interesting question concerns how fast a quantum many body system can thermalize and therefore scramble the quantum information initially encoded in a quantum state.

The Sachdev-Ye-Kitaev model (SYK)~\cite{sachdev93gapless,kitaev,kitaev2018softmode}, describing $N$ Majorana fermions with random two-body all-to-all interactions, has played in this context an important role as minimal model capturing thermalization, scrambling and chaos~\cite{Maldacena_2016,polchinski2016spectrum,Bagrets_2016,Bagrets_2017} or analogously the emergence of strange metals in strongly interacting quantum matter~\cite{sachdev2015bekenstein,chowdhury2021sachdevyekitaev}.

Deformations of the SYK model have been discussed actively. An interesting example is the mixed SYK model, denoted in the following as SYK$_4$+SYK$_2$ model, where an additional random one-body all to all coupling (equivalent to an hopping term) is introduced. This model  in thermal equilibrium has been shown to possess a crossover between a Fermi-Liquid (FL) behavior and a Non-Fermi-Liquid (NFL) regime at low temperatures~\cite{Lunkin_2018,Altland_2019,Lunkin_2020}. This feature, first identified in related models of disordered quantum spins coupled to electrons~\cite{parcollet99nonfermi,georges2001quantum,Cha18341,dumitrescu2021planckian} motivated by the physics of high-Tc superconductors, has recently attracted renewed experimental interest~\cite{husain2020coexisting}.  
At finite $N$ this crossover in the mixed SYK model has been claimed to turn into a transition between chaotic and integrable behavior~\cite{garcia2018chaotic,haque2019eigenstate,micklitz2019nonergodic}, possibly related to many-body localization~\cite{monteiro2021minimal}. 

A large attention has been devoted to the low-energy equilibrium physics of these or related models, to their transport~\cite{song2017strongly,guo2019transport} or scrambling properties as encoded in the growth of out-of-time ordered correlator~\cite{kim2020scrambling}. In particular it was shown that the addition of a relevant perturbation at low energy makes the model less chaotic, with a Liapunov exponent vanishing as a quadratic power-law in temperature~\cite{banerjee2017solvable,guo2019transport,kim2021comment,garcia2021reply}. In this respect, comparatively less work has focused on the genuine nonequilibrium dynamics of SYK models. 

Dynamics in the pure SYK$_4$ model starting from different initial states, including completely uncorrelated ones as well as thermal states of the mixed SYK model, has been studied~\cite{Eberlein_2017,bhattacharya2019quantum}. In the large $N$ limit it was shown that the system thermalizes to an equilibrium state at infinite temperature, unless the initial state is a correlated thermal state of the mixed  SYK$_4$+SYK$_2$ which lead to a finite effective temperature. More recent works have focused on the dynamics of mixed SYK models with complex fermions~\cite{Samui_2021} or deformation of the SYK model possesing a quantum phase transition~\cite{haldar2020quench}. In the high-energy literature the dynamics of pure states in the SYK model have attracted some interest~\cite{kourkoulou2017pure} in particular the dynamics of entanglement entropy~\cite{zhang2020entanglement}. Recently, the periodically driven mixed SYK has been investigated in Ref.~\cite{kuhlenkamp2020periodically}.

In this work we study the nonequilibrium dynamics of the mixed SYK$_4$+SYK$_2$ in the large $N$ limit using Keldysh field theory techniques. Specifically we solve the Kadanoff-Baym equations in real-time and compare their solution with an approach based on a non-perturbative Quantum Boltzmann equation. We start at initial time from the non-interacting SYK$_2$ model and consider different quench protocols, including a sudden switching of the two-body interactions and a double quench of the single particle bandwidth and the interaction. We show that compared to the pure SYK$_4$ the dynamics shows a much richer thermalization landscape, including regimes of fast thermalization and of significant slow-down of heating that we interpret as prethermalization due to proximity to an integrable limit, the pure SYK$_2$ limit. Interestingly we show that a simultaneous quench of single particle bandwidth and interactions leads to a decrease of effective temperature and allows us to explore the crossover from NFL to FL through the quench dynamics and to obtain a nonequilibrium phase diagram for the problem.

The paper is structured as follows. In Sec.~\ref{sec:Model} we introduce the model, the nonequilibrium protocol and present the large-$N$ Kadanoff-Baym equations for its real-time dynamics and discuss how to cast them in the form of a Quantum Boltzmann equation.
 In Sec.~\ref{sec:results1} we present our results for the sudden switching of  random two-body interaction term, while in Sec.~\ref{sec:results2}  we consider the case of a double quench, in which both the interaction and the single particle bandwidth are quenched with respect to the initial condition.  Finally, in Sec.~\ref{sec:discussion} we discuss our results in light of the thermodynamics of the mixed model and show that a double quench makes the effective temperature decrease and heating to slow down and identifies signature of the crossover in the nonequilibrium decay rate evaluated at the effective temperature. In Sec.~\ref{sec:Conclusions} we draw our conclusions, while App.~\ref{app:equilibrium} and App.~\ref{app:QBE} contain further technical details.

\section{The Mixed SYK Model}\label{sec:Model}

We study a generalisation of the SYK model with both quartic and quadratic interactions~\cite{Lunkin_2018,Altland_2019,Lunkin_2020} whose Hamiltonian reads
\begin{equation}
    H(t) = \frac{i}{2} \sum_{i, j = 1}^{N} J_{2, i j}(t) \,  \chi_{i} \chi_{j} -  \, \frac{1}{4 !} \sum_{i, j, k, l = 1}^{N} \, J_{4, i j k l}(t) \chi_{i} \chi_{j} \chi_{k} \chi_{l}
\end{equation}
where $ \chi_{i}$ are Majorana fermions which satisfy $ \left\{ \chi_{i}, \chi_{j} \right\} = \delta_{i j}$. $J_{2,i j}(t)$ and $J_{4, i j k l}(t)$ are time dependent random independent Gaussian variables with zero mean and variances $\overline{J^{2}_{2,i j}(t)} = \frac{J_{2}^{2}(t)}{N}$ and $ \overline{J^{2}_{4,i j k l}(t)} = \frac{3 ! J_{4}^{2}(t)}{N^{3}}$. 

As we will discuss more in detail in Sec.~\ref{sec:results1} we consider as initial condition the ground state of the pure SYK$_{2}$ model and the quench protocol $ J_{2}(t) = \theta(-t) J_{2,i} + \theta(t)  J_{2,f}  $ and $J_{4}(t) = \theta(t) J_{4}$, leaving the possibily to have $ J_{2,i} = J_{2,f} $. In this way we can study both the effect of a pure quench of $J_4$ as well as the combined effect of switching on the interaction and changing the bandwidth of the Majorana fermions.

%
At equilibrium and in the large $N$ limit the SYK model with zero hopping $J_{2} = 0$ describes a non-Fermi liquid where the single particle Green's function in imaginary time decays as $G(\tau) \sim 1 / \sqrt{\tau}$. However this phase is not stable to the introduction of the hopping term $ J_{2} \neq 0$ which constitutes a relevant perturbation and the system turns into a Fermi liquid with single particle Green's function $G(\tau) \sim 1 / \tau$. A cross-over between the Fermi liquid and non-Fermi liquid is expected to happen when the hopping term becomes dominant which corresponds to an energy (or temperature) scale $T^{*} \sim J_{2}^{2} / J_{4}$ \cite{parcollet99nonfermi,song2017strongly}. As we mentioned in the introduction the situation is even more interesting at finite $N$ where the crossover turns into a transition between chaotic and integrable regimes. Here we will not consider the finite $N$ case and focus on the thermodynamic limit and the resulting real-time dynamics that can be also solved exactly through saddle point techniques as we briefly discuss below.

\subsection{Large-N Real-Time Dyson Equation}

The real-time dynamics of the mixed SYK model can be obtained in the large-N limit through saddle point methods on the Keldysh action~\cite{kamenev_2011,Eberlein_2017}. In particular the real-time Green's function of the Majorana fermions, defined as
\begin{equation}
    G^{\alpha \beta}(t_{1},t_{2}) = - \frac{i}{N} \sum_{i} \langle \chi_{i}^{\alpha}(t_{1}) \chi_{i}^{\beta}(t_{2})\rangle
\end{equation}
with $\alpha,\beta=\pm $ Keldysh contour index, can be shown to satisfy a real-time Dyson Equation of the form
\begin{equation}\label{eqn:Dyson}
    \Big[  \hat{G}_{0}^{-1} - \hat{\Sigma} \Big] \circ \hat{G} = 1,
\end{equation}
where  $[\hat{G}_{0}^{-1}]^{\alpha \beta}(t_{1},t_{2}) = i \alpha \delta_{\alpha \beta} \delta(t_{1}-t_{2}) \partial_{t_{1}}$ is the free Majorana fermions Green's function, the self-energy reads
\begin{equation}\label{eq:sigmaalphabeta}
\begin{split}
    \Sigma^{\alpha \beta}(t_{1},t_{2}) &= - \alpha \beta J_{4}(t_{1}) J_{4}(t_{2}) G^{\alpha \beta}(t_{1},t_{2})^{3} \\
    &+J_{2}(t_{1}) J_{2}(t_{2}) G^{\alpha \beta}(t_{1},t_{2})
\end{split}
\end{equation}
and the symbol $\circ $ denotes real-time convolution. For Majorana fermions it is convenient to work with the greater (lesser) Green function $G^{>(<)}(t_{1},t_{2})$ which are related as
%
%
\begin{equation}
    G^{>}(t_{1},t_{2}) = - G^{<}(t_{2},t_{1})
\end{equation}
and from which one can obtain all the relevant Green's functions, 
%

\begin{align}
    G^{R}(t_{1},t_{2}) &\equiv \theta(t_{1} - t_{2}) \big( G^{>}(t_{1},t_{2}) - G^{<}(t_{1},t_{2}) \big) \\
    G^{A}(t_{1},t_{2}) &\equiv - \theta(t_{2} - t_{1}) \big( G^{>}(t_{1},t_{2}) - G^{<}(t_{1},t_{2}) \big) \\
    G^{K}(t_{1},t_{2}) &\equiv  G^{>}(t_{1},t_{2}) + G^{<}(t_{1},t_{2})
\end{align}
The first Schwinger-Dyson equation can be put into a more convenient form known as the Kadanoff-Baym (KB) equations~\cite{kadanoffbaym}:
\begin{align}\label{eqn:kb1}
    i \partial_{t_{1}} G^{>,<}(t_{1},t_{2}) &= \int_{- \infty}^{+ \infty} d t  \Sigma^{R}(t_{1},t) G^{>,<}(t,t_{2}) + \nonumber\\
   &+ \int_{- \infty}^{+ \infty} d t  \Sigma^{>,<}(t_{1},t) G^{A}(t,t_{2})  \\\label{eqn:kb2}
          - i \partial_{t_{2}} G^{>,<}(t_{1},t_{2}) &= \int_{- \infty}^{+ \infty} d t  G^{R}(t_{1},t) \Sigma^{>,<}(t,t_{2}) +\nonumber\\
& G^{>,<}(t_{1},t)\Sigma^{A}(t,t_{2})  
\end{align}

Likewise we define the retarded and advanced self-energies:

\begin{align}\label{eq:sigmaRA}
    \Sigma^{R}(t_{1},t_{2}) &\equiv \theta(t_{1} - t_{2}) \big( \Sigma^{>}(t_{1},t_{2}) - \Sigma^{<}(t_{1},t_{2}) \big) \\
    \Sigma^{A}(t_{1},t_{2}) &\equiv - \theta(t_{2} - t_{1}) \big( \Sigma^{>}(t_{1},t_{2}) - \Sigma^{<}(t_{1},t_{2}) \big) 
\end{align}

\begin{figure}[t]
	\includegraphics[width=0.45\textwidth]{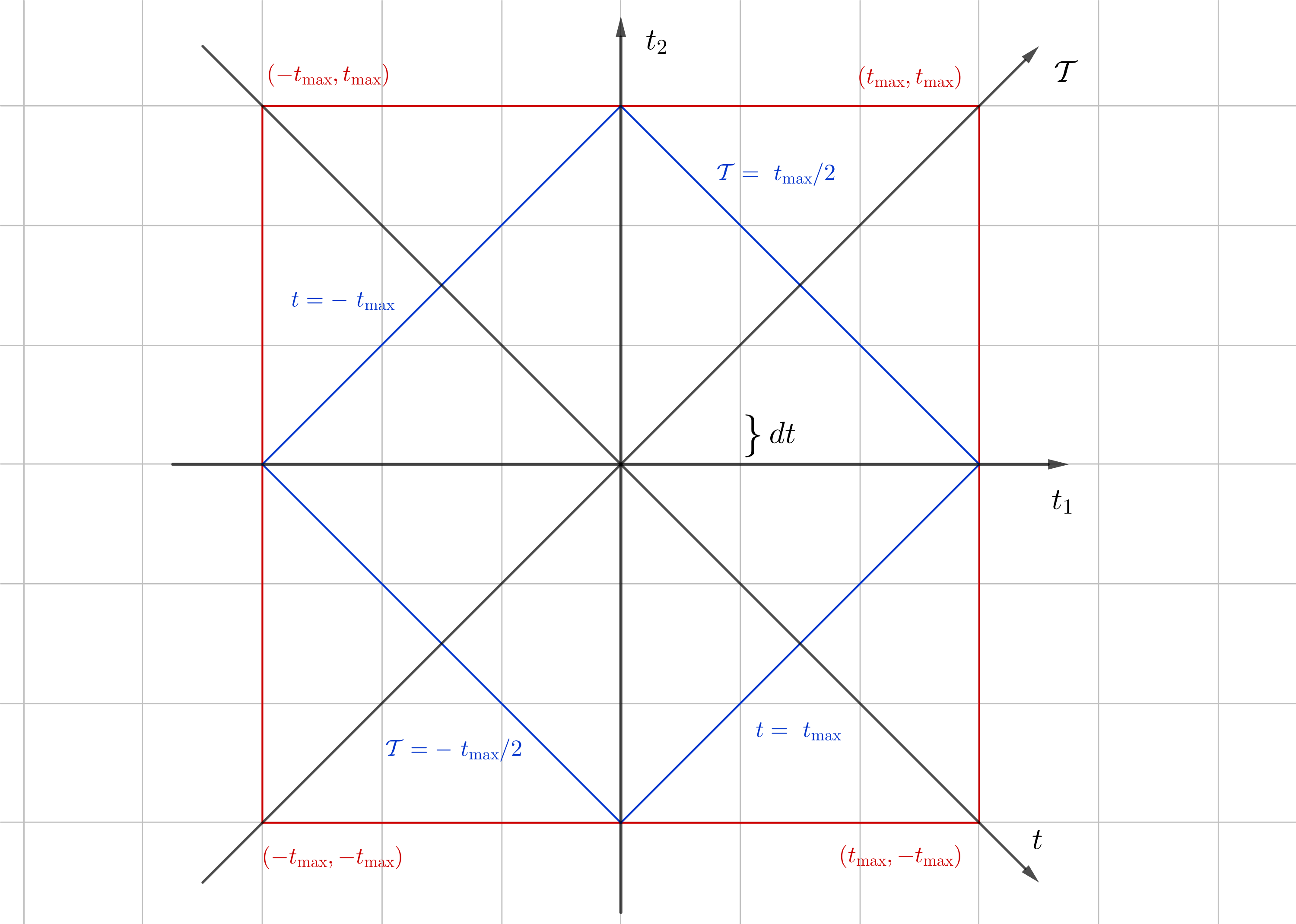}
	\caption{Sketch of the $t_1,t_2$ plane as well as the rotated plane $\mathcal{T},t$.}
	\label{fig1}
\end{figure}

We mention here for future use that the greater Green's function of the pure SYK$_{2}$ model ($J_{2} (t) = J_{2}$, $J_{4}(t) = 0$) is known exactly, at least in the long-time stationary (i.e. time-translational-invariant limit)

\begin{equation}\label{eqn:GSYK2}
\begin{split}
    G^{>}_{2}(t) &= \frac{- i}{2 J_{2} t} \left( J_{1}[ 2 J_{2}t]
    - i H_{1}[2 J_{2} t ] \right)
\end{split}
\end{equation}
where $J_1(x),H_1(x)$ are special (Bessel's and Henkel's) functions of the first kind.

\subsubsection{Numerical Integration of Kadanoff-Baym Equations}

%


In this paper we solve the KB equations~(\ref{eqn:kb1}-\ref{eqn:kb2}) numerically on a $t_{1} - t_{2}$ grid of size $2001\times2001$, $4001\times4001$ or $6001\times6001$ with time step $d t = 0.05$. The grid has a length $2 \, t_{\textrm{max}}$ in each direction as shown in red in Fig.~\ref{fig1}. Initially at times $t_{1},t_{2} < 0$ the system is prepared in the ground state of the pure SYK$_{2}$ model with coupling $J_{2} = 0.5$. We solve KB equations for $G^{>}$ on the grid by using a predictor-corrector scheme \cite{Eberlein_2017} \cite{bhattacharya2019quantum}, \cite{haldar2020quench}. Integrals are computed using the trapezoidal rule. We verify the consistency of our numerical code by checking the conservation of energy and the normalization of the spectral density. Thermal equilibrium solutions for a given inverse temperature $\beta$ are obtained by solving the Schwinger-Dyson equation self-consistently following \cite{Eberlein_2017}, further details are discussed in Appendix~\ref{app:equilibrium}.

\subsection{Observable of Interests: Spectral and Distribution Functions}

In order to interpret the results obtained by the numerical integration of the real-time KB equations  it is convenient to introduce a  mixed time-frequency representation for the Green's functions defined in the previous section. Specifically we can change the time coordinates $(t_{1},t_{2})$ to an average and relative time coordinates $( \mathcal{T}, t)$ defined as
\begin{equation}
    \mathcal{T} = \frac{t_{1} + t_{2}}{2}, \quad t = t_{1} - t_{2}
\end{equation}
The bounds of the $ (\mathcal{T},t)$ grid are shown in blue in Fig.~\ref{fig1}. Notice that on this grid the initial conditions corresponds to $\mathcal{T} = - \mathcal{T}_{\textrm{max}}$  and the maximum $\mathcal{T}$ value to $ \mathcal{T} = \mathcal{T}_{\textrm{max}}  $ with $ \mathcal{T}_{\textrm{max}} = t_{\textrm{max}} / 2 $.

Taking the Fourier transform with respect to the relative time $t$ we define the so-called Wigner transform of the Green's functions~\cite{kamenev_2011,Eberlein_2017} 
\begin{equation}
    G(\mathcal{T}, \omega) = \int \, d t \, e^{i \omega t} G \Big( t_{1} = \mathcal{T} + \frac{t}{2}, t_{2} = \mathcal{T} - \frac{t}{2}  \Big)
\end{equation}
This has the advantage of showing explicitly the effect of time-translation symmetry breaking due to the quench and resulting in an explicit dependence on the average time $\mathcal{T}$. Furthermore it suggests a picture of slow-varying quasi-equibrium which connects naturally with the long-time limit in which one expects the approach to thermal equilibrium.

Using the Wigner transform we can define quantities which have a direct physical interpretation such as the time-dependent spectral function $A(\mathcal{T},\omega)$
\begin{equation}
    A(\mathcal{T},\omega) = - 2 \textrm{Im} G^{R}(\mathcal{T},\omega)
\end{equation}
or distribution function $F(\mathcal{T},\omega)$
\begin{equation}\label{eq:Fomega}
    iG^{K}(\mathcal{T},\omega) = F(\mathcal{T},\omega) A(\mathcal{T},\omega)
\end{equation}
using a parametrization of the Keldysh Green's function which evokes explicitly a fluctuation-dissipation theorem. In fact in thermal equilibrium, corresponding  to the initial condition or possibly the long-time behavior if thermalization is established, those two quantities are time-independent and related by a universal identity, the fluctuation-dissipation theorem~\cite{altland_simons_2010}
\begin{equation}\label{eq:fdt}
    iG^{K}(\omega) = \tanh \Big( \frac{\beta \omega}{2} \Big) A(\omega)
\end{equation}
where $\beta$ is the inverse temperature. 

\subsection{Quantum Boltzmann Equation Approach}\label{sec:qbe}

In this section we discuss a different way of solving the Kadanoff-Baym Equations~(\ref{eqn:kb1}-\ref{eqn:kb2}) which is based on a coarse-grained approach in the time-domain, focusing on the long-time dynamics rather than on the short-time transient. This idea can be formalized by performing a Wigner transform of the KB equations followed by a gradient expansion in the \emph{slow} average time $\mathcal{T}$, which casts the KB equation into a quantum kinetic (Boltzmann) equation (QBE)~\cite{kamenev_2011}  for the Wigner-transformed distribution $F(\mathcal{T},\omega)$. Performing these steps for our mixed SYK model we obtain (as we discuss in detail in Appendix~\ref{app:QBE}), to the lowest order in the gradient expansion, the equation
\begin{equation}
    \partial_{\mathcal{T}} F(\mathcal{T},\omega) = I^{\textrm{coll}} \big[ F(\mathcal{T},\omega) \big]
\end{equation}
where the collision integral $I^{\textrm{coll}} $ reads
\begin{equation}\label{eqn:Icoll}
\begin{split}
I^{\textrm{coll}} \big[ F(\mathcal{T},\omega) \big] = \, &i \Sigma^{K}(\mathcal{T},\omega) - i F(\mathcal{T},\omega) \times \\
&\big( \Sigma^{R}(\mathcal{T},\omega) - \Sigma^{A}(\mathcal{T},\omega) \big)
\end{split}
\end{equation}
and we emphasize that the self-energy components  $\Sigma^{R}(\mathcal{T},\omega)$, $\Sigma^{A}(\mathcal{T},\omega)$ and $\Sigma^{K}(\mathcal{T},\omega)$, which can be obtained from Eq.~(\ref{eq:sigmaalphabeta}) after Wigner transform, also depend on the distribution function itself (see Appendix~\ref{app:QBE} for their explicit expression). The dynamics of the distribution function $F(\mathcal{T},\omega)$ is in principle coupled, through the full KB equations, to the dynamics of the spectral function $A(\mathcal{T},\omega)$ which also enters the self-energies in Eq.~(\ref{eqn:Icoll}). However, to the lowest order in the gradient expansion one can show that the dynamics of the spectral function is simply given by
\begin{equation}\label{eqn:partialTA}
    \partial_{\mathcal{T}} A(\mathcal{T},\omega) = 0\,,
\end{equation}
i.e. the spectral function of the system is already stationary while the distribution function is still slowly evolving. As we are going to see in the next section this separation of time scales between spectrum and occupation is to a good extent present in the full solution of the KB equations, thus supporting the validity of the lowest order gradient expansion at least for what concerns the long-time dynamics.
Before closing it is worth emphasizing that traditionally QBE approaches have been used together with perturbative evaluations of the collision integral and under specific assumptions on the spectral properties of the systems (quasiparticle approximation). Here instead our collision integral in Eq.~(\ref{eqn:Icoll}) is \emph{exact} due to the large-$N$ nature of the SYK problem. We note that recently similar QBE approaches have been developed to describe the real-time dynamics of other strongly interacting quantum many-body systems beyond perturbation theory~\cite{lindner2006comparison,tavora2013quench,picano2021quantum}.

\section{Results: Quench of $J_4$}
\label{sec:results1}

In this section we present our results for the dynamics of the mixed SYK model as obtained  from the solution of the real-time Dyson equation. Specifically we consider first a sudden switching of the quartic interaction $ J_{4}(t) = \theta(t) J_{4} $, starting from a pure SYK$_2$ model in its ground state and with $J_2=0.5$. Later in Sec.~\ref{sec:results2} we will consider the effect of quenching both $J_2$ and $J_4$.

\subsection{Transient Spectral Function and Thermalisation of the mixed SYK model}

\begin{figure}[t!]
	\includegraphics[width=0.45\textwidth]{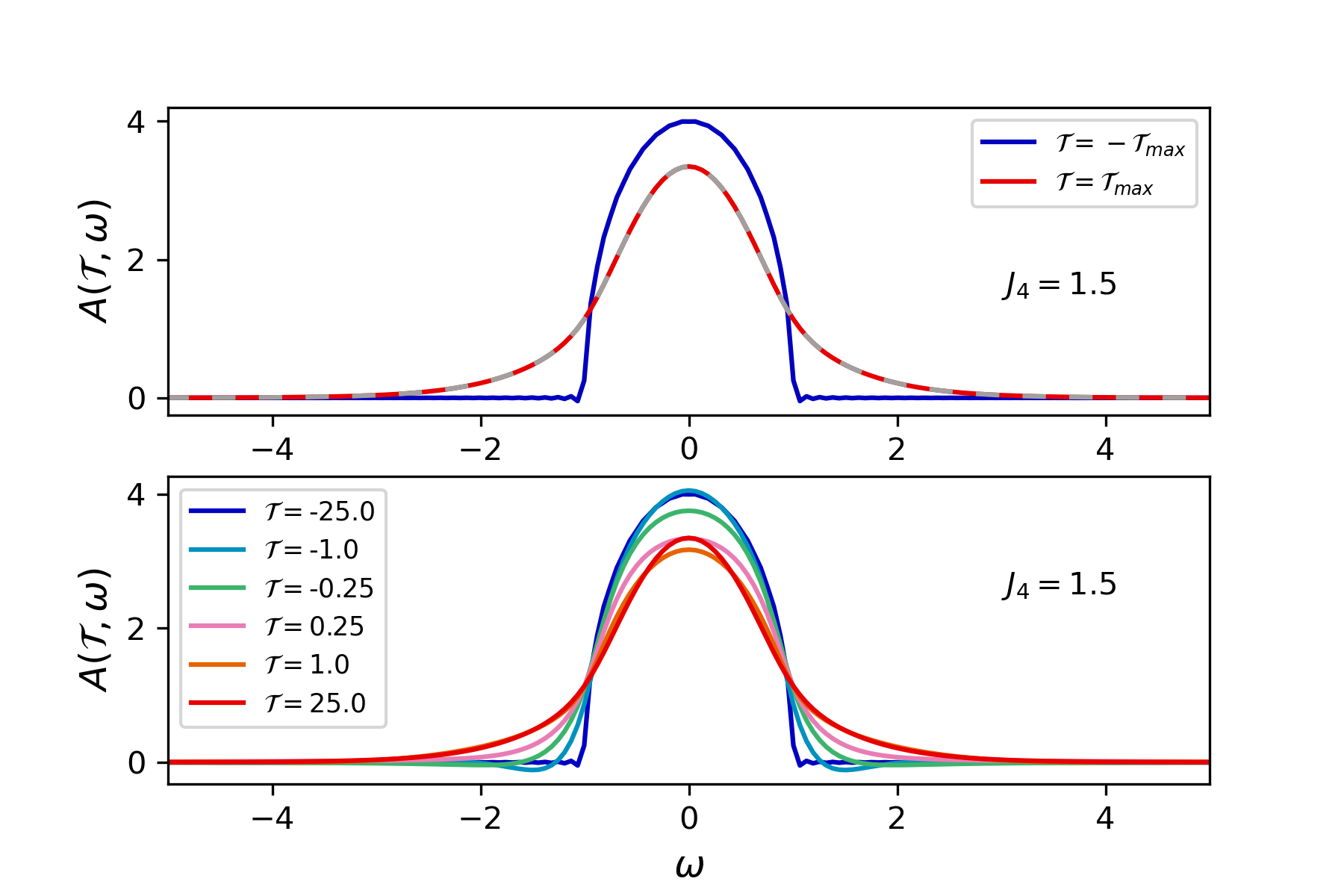}
	\caption{Transient Spectral Function after a sudden switch-on of $J_4$ in the mixed SYK model. 
	Top: Spectral function at initial time ($\mathcal{T}=-\mathcal{T}_{\rm max}$) and at long times ($\mathcal{T}=\mathcal{T}_{\rm max}$), compared to the equilibrium spectral function of the mixed SYK$_{4}$+SYK$_{2}$ model at the final temperature (dashed grey line), confirming thermalisation and evolution of the spectral function for different times $\mathcal{T}$. 	}\label{fig2}
\end{figure}

We start discussing the evolution of the spectral function $A(\mathcal{T},\omega)$ after a quench to $J_4=1.5$. In Fig.~\ref{fig2} we plot the initial spectral function of the SYK$_2$ model, which features the well known sharp edge semicircular density of state, and its long-time limit after the switching on of the $J_4$ interaction, featuring a much broader resonance with tails at higher frequencies. We compare the latter with the equilibirum spectral function of the mixed SYK model evaluated at the final temperature (see next section)  and find a perfect match thus confirming that the mixed SYK model reaches thermal equilibrium. In the bottom panel we plot the evolution of the spectral function $A(\mathcal{T},\omega)$  as a function of increasing time $\mathcal{T}$. We see that spectral features at high frequencies reshape rather rapidly after the quench, with tails forming above the bandwidth set by $J_2$, while the low frequency features follow at later times.

\begin{figure}[t!]
	\includegraphics[width=0.45\textwidth]{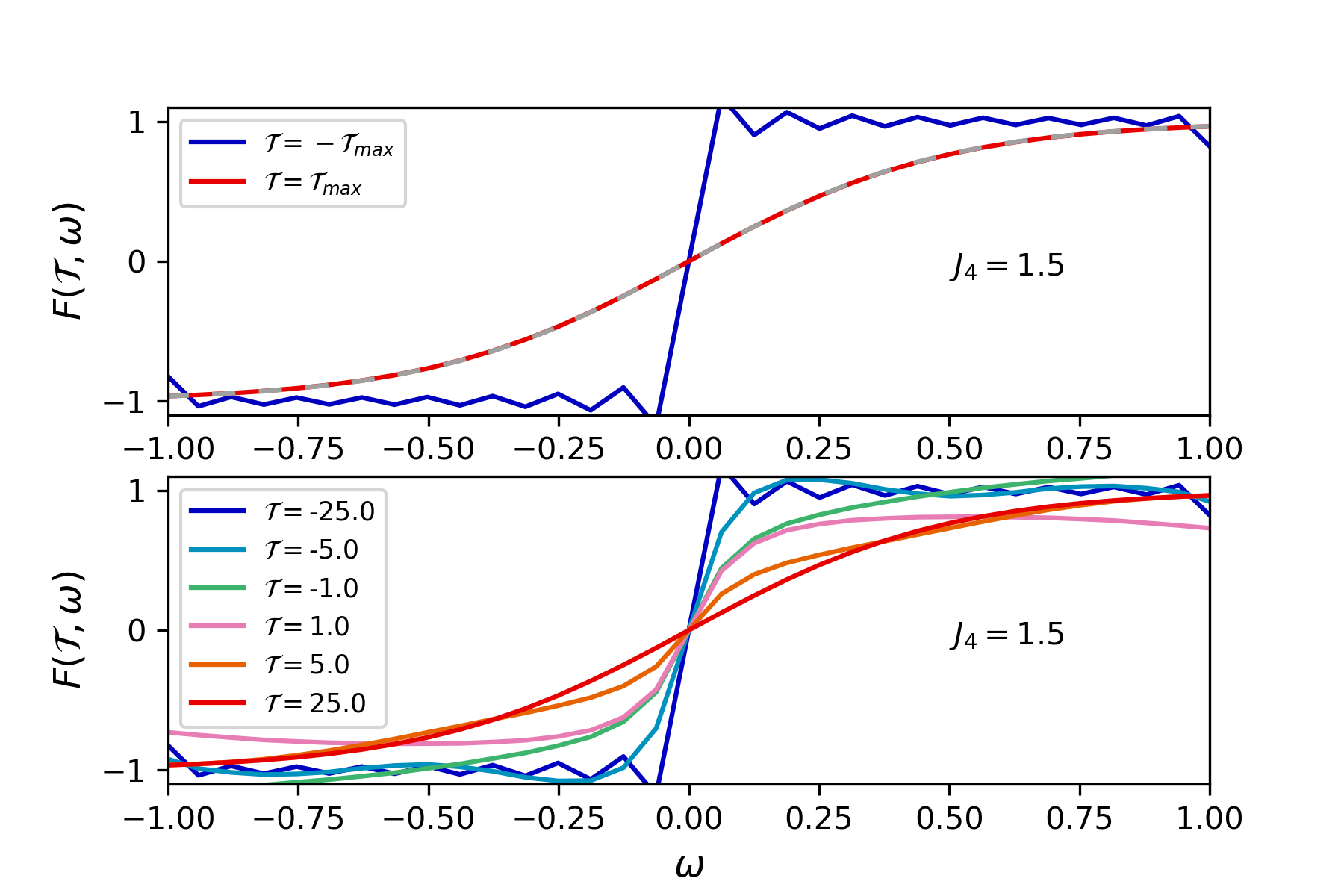}	
	\caption{Transient Distribution Function after a sudden switch-on of $J_4$ in the mixed SYK model. 
	Top: Distribution function at initial and final times and its dynamical evolution showing the onset of heating (finite temperature). Parameters: $J_2=0.5$, $J_4=1.5$.
	}\label{fig3}
\end{figure}
\subsection{Distribution Function and Dynamics of Effective Temperature}

A further demonstration of the onset of thermalization in the mixed SYK model comes from the dynamics of the distribution function, $F(\mathcal{T},\omega)$, defined in Eq.~\ref{eq:Fomega}, that we plot in Fig.~\ref{fig3}. As previously, we show in the top panel the initial condition and the long-time limit, while in the bottom panel the evolution of the distribution function for different average times $\mathcal{T}$.
As the initial state is taken at zero temperature, the distribution function is rather sharp around low frequency, while long time after the quench a smoother behavior is approached indicating the development of a higher final temperature. From the bottom panel we can see that, as for the spectral function, the high-frequency features of the distribution are those that re-adjust the faster after the quench, while the low frequency sector takes longer time to respond. Furthermore, by comparing Fig.~\ref{fig2}-\ref{fig3}, one can conclude that the dynamics of the spectral function is typically faster than the one of the distribution function.

\begin{figure}[t]
	\includegraphics[width=0.45\textwidth]{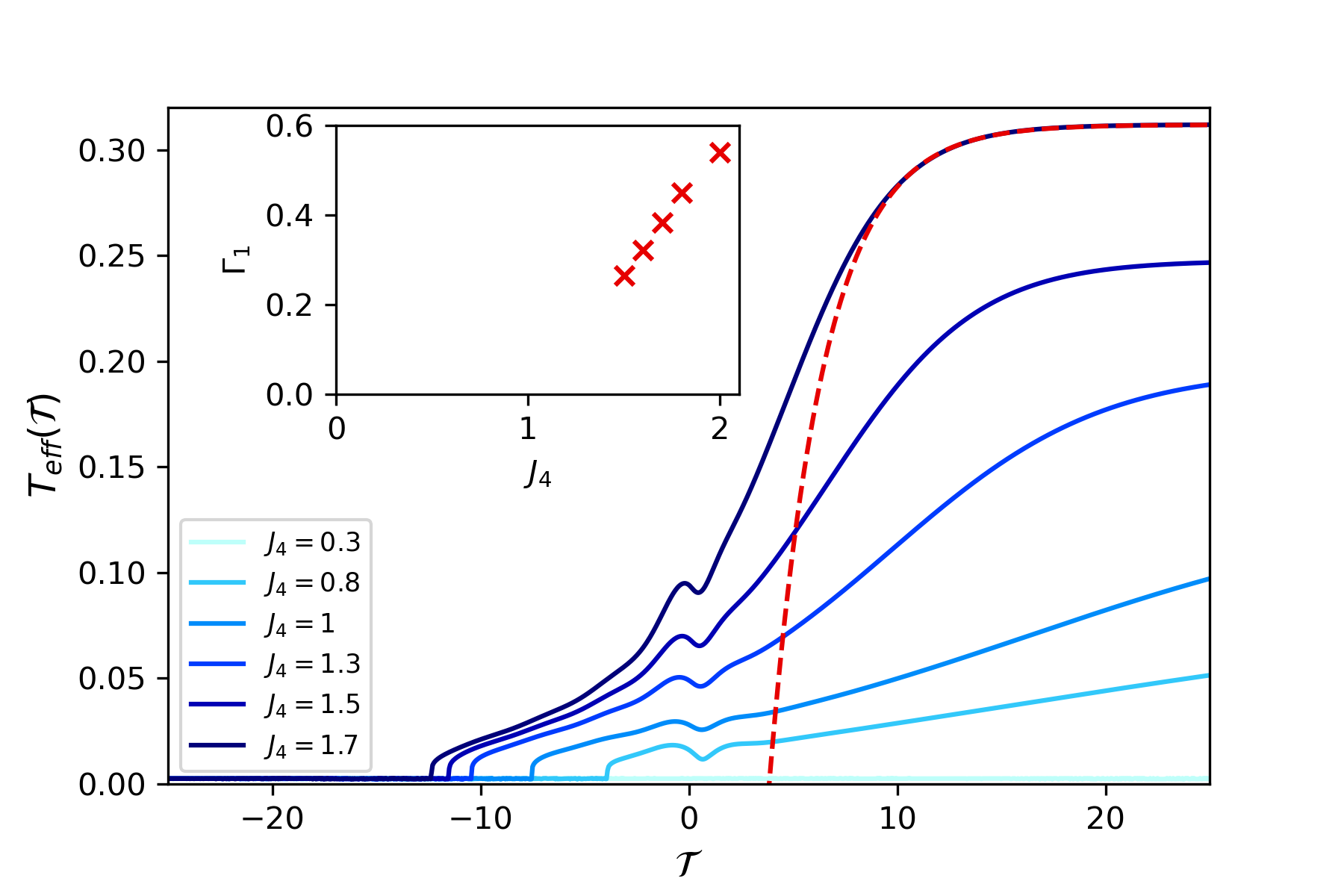}
	\caption{Dynamics of the effective temperature $T_{\rm eff}(\mathcal{T})$, extracted from the distribution function as discussed in the main text, for different values of the $J_4$ interaction.  We see that the heating dynamics of the system strongly depends on the value of $J_4$. }\label{fig4}
\end{figure}

To better understand the onset of thermalization we note that throughout the evolution the low frequency behavior of the distribution function is linear in frequency, which suggests to extract a time-dependent effective temperature $ T_{\textrm{eff}}(\mathcal{T})$ by fitting the low frequency behavior of $F(\mathcal{T},\omega)$ with $ \tanh ( \beta_{\textrm{eff}}(\mathcal{T}) \omega / 2 )$. We plot the dynamics of this effective temperature for different values of $J_4$ in Fig.~\ref{fig4}. We see that for small quenches the system remains close to the ground-state and the effective temperature changes slowly with time. One can understand this slowing down of heating in the weak quench regime in terms of a prethermalization phenomenon~\cite{berges2004prethermalization,moeckel2008interaction,langen2016prethermalization}, found in other quench problems of interacting quantum many-body systems and associated to the proximity to an integrable point, in the present case the pure SYK$_2$ model.

On the other hand increasing the strength of the quench $J_4$ leads to a faster heating dynamics. The approach to the long-time limit, that we identify with the final temperature reached by the system after the quench,
$T_f=T_{\textrm{eff}}(\infty)$, is exponential
$$
T_{\textrm{eff}}(\mathcal{T}) - T_{f} = A \exp\left(-\Gamma_1 \mathcal{T}\right)
$$
from which we can extract a thermalization rate $\Gamma_1$ that depends in general on the quench parameters, in particular on the final value of $J_4$, as we show in the inset. In particular for quenches of intermediate strength we find $\Gamma_1\sim J_4$, while upon decreasing we expect a subleading behavior, possibly quadratic in $J_4$. However the time-scales needed to reach equilibrium exceed our $\mathcal{T}_{\rm max}$ therefore we cannot conclude on the nature of this scaling (see however next section).

\begin{figure}[t]
	\includegraphics[width=0.45\textwidth]{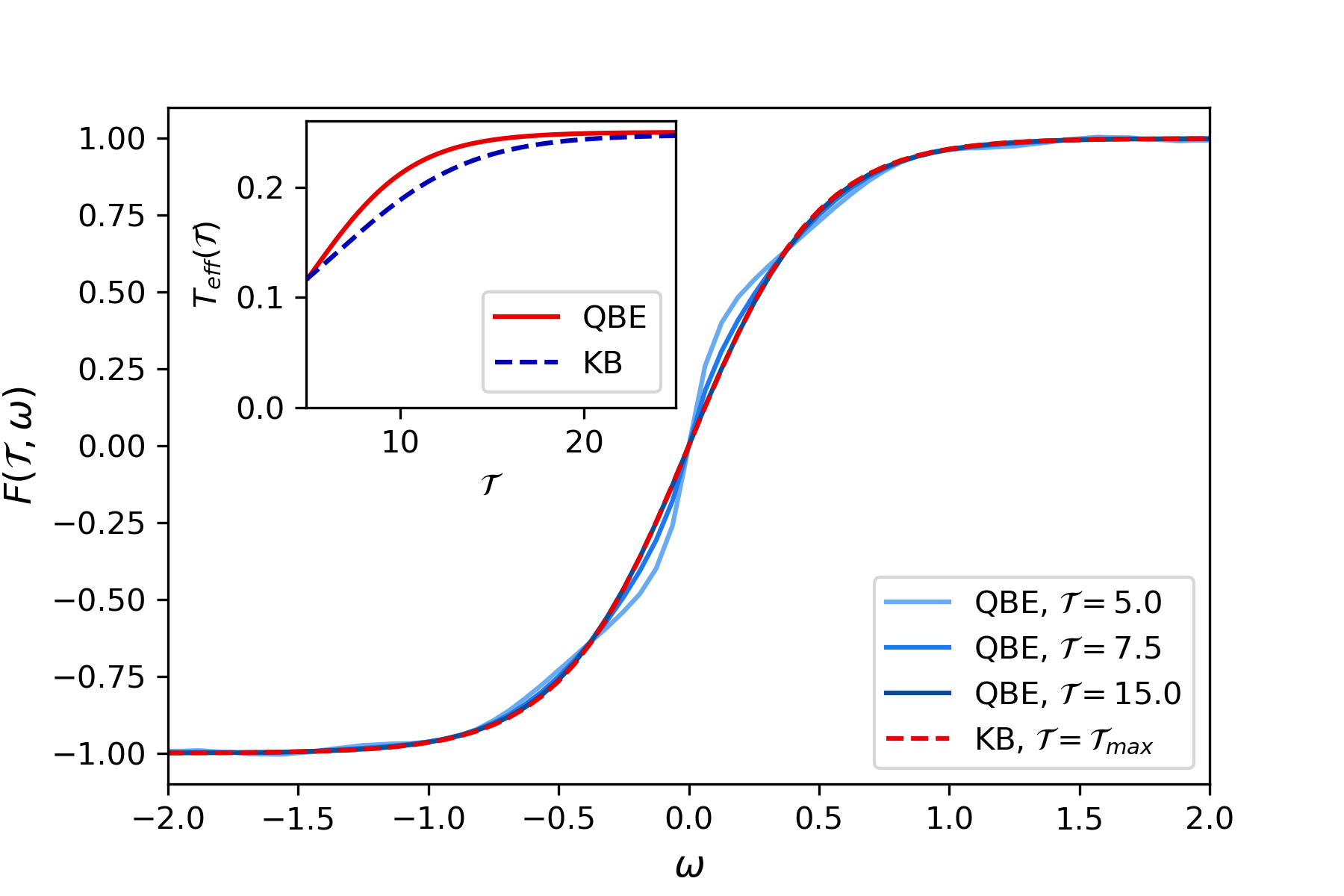}
	\caption{
Evolution of the distribution function $F(\mathcal{T},\omega)$ at different times compared with the long-time limit obtained from the full solution of the KB equations. Inset: Evolution of the time-dependent effective temperature obtained by fitting the QBE distribution function at low-frequency, compared with the one obtained from KB equations.
	Quench Parameters:$( J_{2,i} = 0.5,0 ) \rightarrow (J_{2,f} = J_{2,i}, J_{4} = 1.5$)	} 
	\label{fig5new}
\end{figure}

\begin{figure*}[t!]
\begin{center}
  \subfigure[]{ \includegraphics[scale = 0.45]{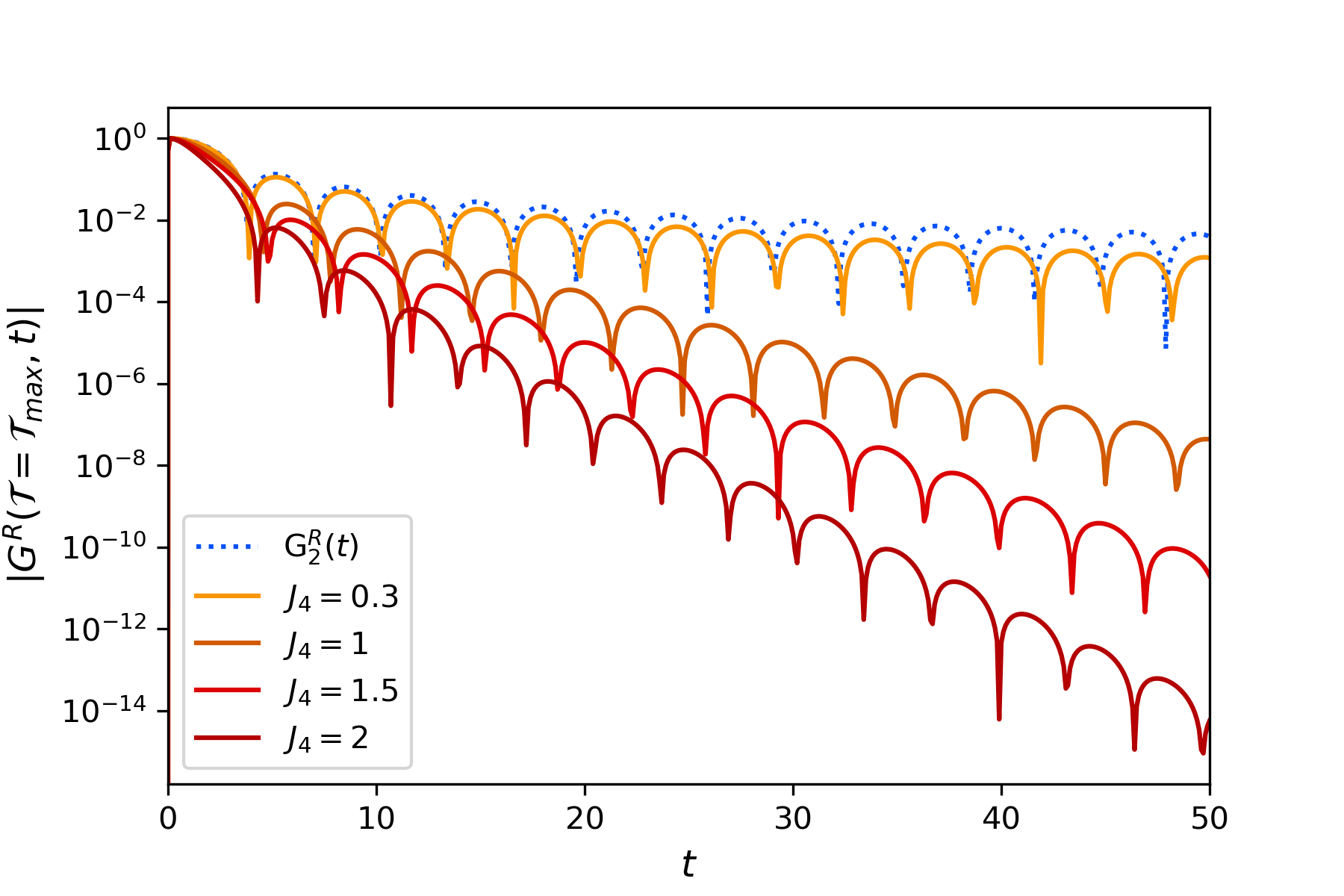}}
  \quad
  \subfigure[]{\includegraphics[scale = 0.45]{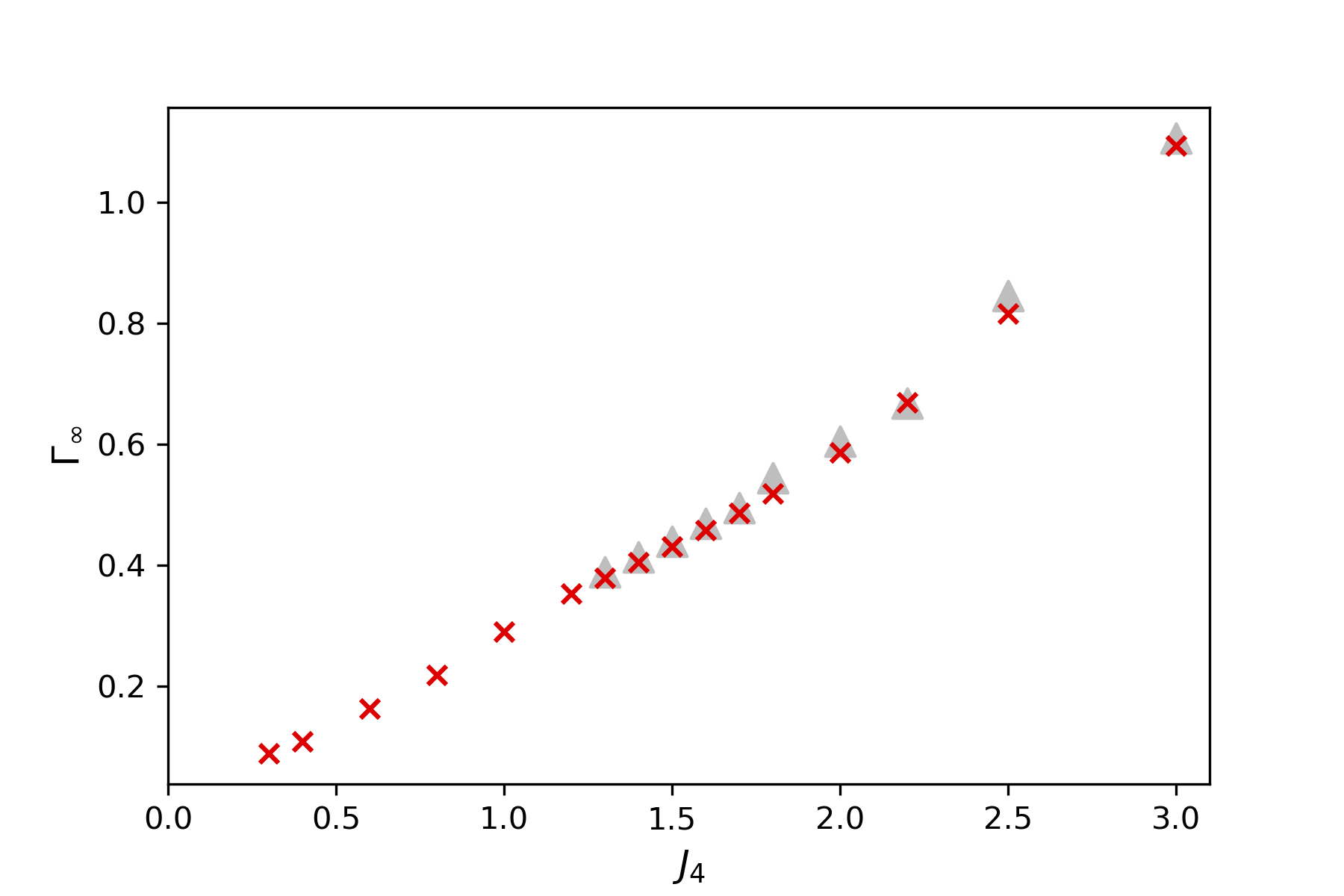}}
  \caption{(a) Real-time retarded Green's function as a function of the relative time $t$ long after the quench ($\mathcal{T} = 25$) for different $J_{4}$. We compare the decay with the power-law behavior expected for a pure SYK$_2$ (dotted blue line). (b) Exponential decay rate $ \Gamma_{\infty}$ extracted from the large $\mathcal{T}$ limit of the retarded Green function for different $J_{4}$ (red crosses). The grey triangles are the decay rates extracted from the equilibrium Green's function at $T_{f}$.  }
  \label{fig5}
\end{center}
\end{figure*}
\subsection{Comparison With QBE Approach}

We now compare the dynamics obtained from the full solution of the KB equations with the QBE approach discussed in Sec.~(\ref{sec:qbe}).  To this extent we focus on a sudden switching of the $J_4$ interaction, keeping the $J_2$ coupling fixed.
An important issue when performing this comparison concerns the choice of initial conditions for the distribution and spectral function entering the QBE approach. As we have seen the spectral function remains constant in time at the lowest order approximation (see Eq.~(\ref{eqn:partialTA})), thus we take it to be equal to the equilibrium spectral function of the final Hamiltonian at the final effective temperature.  The idea behind this choice is that, as the full KB solution shows, the spectral function thermalizes on much faster time scales than the distribution function. For what concerns the distribution function the issue is more subtle since any equilibrium thermal like distribution annihilates the collision integral, independently on the value of the temperature (see Appendix \ref{app:QBE}). We therefore choose to take as our initial condition the distribution function obtained by solving the full KB equations for a short time interval $\mathcal{T}_{0} = 5$.
In Fig.~(\ref{fig5new}) we plot, for the quench $( J_{2,i} = 0.5,0 ) \rightarrow (J_{2,f} = J_{2,i}, J_{4} = 1.5$), the evolution in time of the QBE distribution function from the initial $\mathcal{T}=\mathcal{T}_{0} = 5$ to late times, and show that it correctly relaxes to the thermally equilibrated one (red-dashed lines) obtained from the KB dynamics. In the inset, we compare the dynamics of the effective temperature extracted with the two methods. We confirm that the discrepancy between QBE and KB dynamics is mainly in the intermediate transient while the long-time asymptotics is well captured. 

\subsection{Real-time Retarded Green's functions: Long-time Decay Rate and Waiting Time dependence}

We now discuss the dynamics of the retarded Green's function in the time-domain, focusing in particular on its long time decay. From the analysis of the spectral function presented in the previous section it appeared difficult to discuss the effect of the quench on the low frequency behavior of the spectrum, a feature that will appear more clearly in the time domain.
In the left panel of Fig.~\ref{fig5} we plot the retarded Green's function $G^R(\mathcal{T},t)$ long time after the quench, i.e. for $\mathcal{T}=\mathcal{T}_{\rm max},$  as a function of the relative time $t$ and for different values of $J_4$.  

We first note that the Green's function shows pronounced oscillations, with a period which appears to be independent of $J_4$ and completely set by the $J_2$ scale. This can be understood by considering the analytic expression for the Green's function of the pure SYK$_2$ model given in Eq.~\ref{eqn:GSYK2}, which indeed show oscillations with a frequency set by $J_2$. 

Concerning the long-time decay we see that with the exception of the very small quench regime, where the Green's function remains closer to the one of a pure SYK$_{2}$ which decays as a power law,  the system rapidly enters into an exponential decay regime as a function of the relative time, which is compatible with the Lorentzian lineshape found in the frequency-resolved spectral function (see Fig.~\ref{fig2}). We can therefore extract a decay rate for the stationary retarded Green's function using an ansatz of the form
\begin{equation}\label{eqn:Gammainf}
G^R(\mathcal{T}_{\rm max},t)\sim \exp(-\Gamma_{\infty} t)\,
\end{equation}

\begin{figure}[t]
	\includegraphics[width=0.45\textwidth]{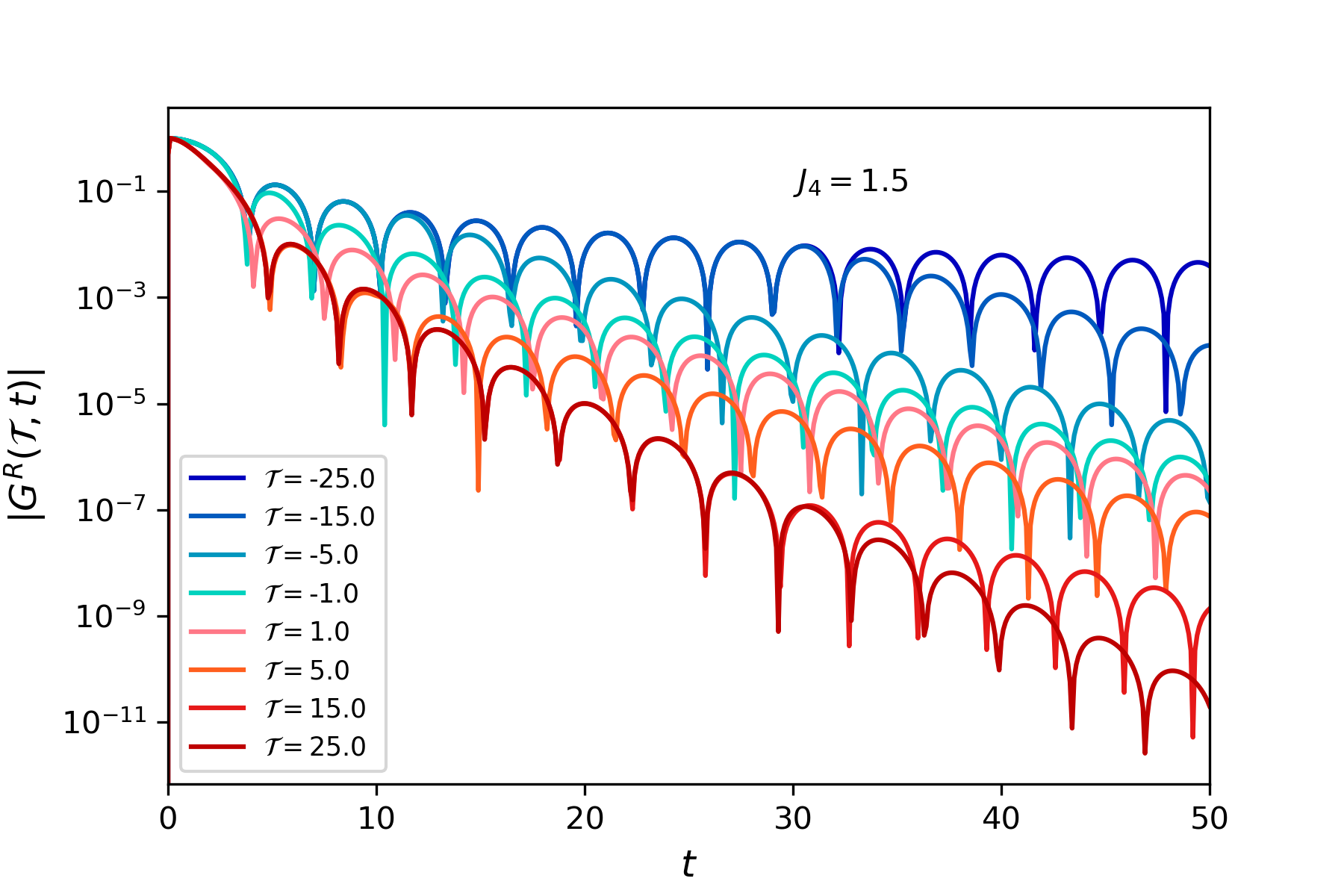}
	\caption{Waiting time dependence of the real-time retarded Green's function after a quench of $J_4=1.5$. At short average time the decay as a function of the relative time $t$ is power law, reflecting the initial low temperature state of the SYK$_2$ model. Upon increasing the time after the quench $\mathcal{T}$ the decay crosses over to an exponential decay, around a time $t\sim 2\mathcal{T}$. We note that the decay at long time is exponential, yet with a decay rate which is slower than the long-time limit $\Gamma_{\infty}$. This behavior is compatible with an ansatz of the form~\ref{eqn:ansatz} for the two-times Green's function.}\label{fig6}
\end{figure}

As for the thermalization time discussed before, the decay rate $\Gamma_{\infty} $ depends in general on the quench parameters, in particular the value of the interaction $J_4$, as we plot in the right panel of Fig.~\ref{fig5}.  is very small for $J_4\rightarrow0$ reflecting the power-law decay of the pure SYK$_2$ model and grows upon increasing the strength of the quench showing a linear behavior at intermediate couplings and a superlinear regime for large $J_4$. Finally we expect (not shown) the decay rate to saturate at larger values of $J_4$. 

%

We now move on to discuss the waiting time dependence of the real-time retarded Green's function which further allows to clarify how the initial power law decay regime and the long-time exponential decay regime are connected as the average time $\mathcal{T}$ is increased.
In Fig.~\ref{fig6}  we plot the real-time  retarded Green's function, $G^R(\mathcal{T},t)$ as a function of the relative time, for different average times $\mathcal{T}$ and for a quench to $J_4=1.5$. We recognize a characteristic feature of this two-time Green's function, namely a short time behavior and a long time regimes with a crossover at times $ t\sim 2\mathcal{T}$. We also notice that the short-time behavior is power law for negative average times $\mathcal{T}<0$, corresponding to the regime of influence of the initial condition, while it is exponential for positive average times $\mathcal{T}>0$. The numerical results suggest an ansatz of the form
\begin{align}\label{eqn:ansatz}
G^R(\mathcal{T},t)=G^R_{\infty}(t)f\left(\mathcal{T},t\right)
\end{align}
where $f\left(\mathcal{T},t\right)$ is a function chosen to capture the crossover and the short/long time behavior of the two-times retarded Green's function~\cite{schiro2014transient}. In particular for positive times $\mathcal{T}>0$ we have that $f\simeq 1$ for $t\ll 2\mathcal{T}$ while it grows exponentially for long times, i.e.
$f\sim \exp(\Gamma' t)$ for $t\gg 2\mathcal{T}$, with a rate $\Gamma'$ which is in general smaller than the long-time asymptotic rate $\Gamma_{\infty}$, leading to a slow down of the decay rate at finite positive waiting times, as shown in Fig.~\ref{fig6}.

\section{Results: Quench of $J_4$ and $J_2$}\label{sec:results2}

The results of previous section have shown that under a sudden switching of the interaction the mixed SYK model thermalizes rapidly, with a significant increase of the effective temperature, except for weak quenches $J_4\ll J_2$ where a much slower heating dynamics is observed that we have interpreted as a signature of prethermalization.
In order to explore further the thermalization pathways of the mixed SYK model and to connect them to its equilibrium properties, in particular to the Non-Fermi-Liquid to Fermi Liquid crossover, in this section we consider a double quench protocol where in addition to the sudden switching of the $J_4$ interaction the system at time $t=0$ undergoes also a quench of the single particle term $J_{2,i}=0.5 \rightarrow J_{2,f} \neq J_{2,i}$.  Protocols of this sort have been discussed before in the literature on quantum quenches, see for example Refs.~\cite{mitra2011mode,schiro2014transient,mitra2018quantum} with the idea of disentangling the role of integrability-breaking and generic nonequilibrium perturbations in the heating production and thermalization pathways of quantum systems. We emphasize that while the equilibrium properties of the mixed SYK model are completely controlled by the adimensional ratio $J_4/J_2$, for the non-equilibrium dynamics there are in principle two independent adimensional quantities, $J_{2,f}/J_{2,i}$ and $J_4/J_{2,f}$, the former controlling the degree of excitation of single particle states while the latter the role of interactions. One can naively expect that when single particle excitations dominate over interactions, scattering processes are kinematically blocked resulting in a slow down of heating. As we are going to show below, this double quench protocol provides us indeed with an additional tuning parameter to explore the low-heating regime of the mixed SYK quench dynamics.

\begin{figure}[t!]
	\includegraphics[width=0.45\textwidth]{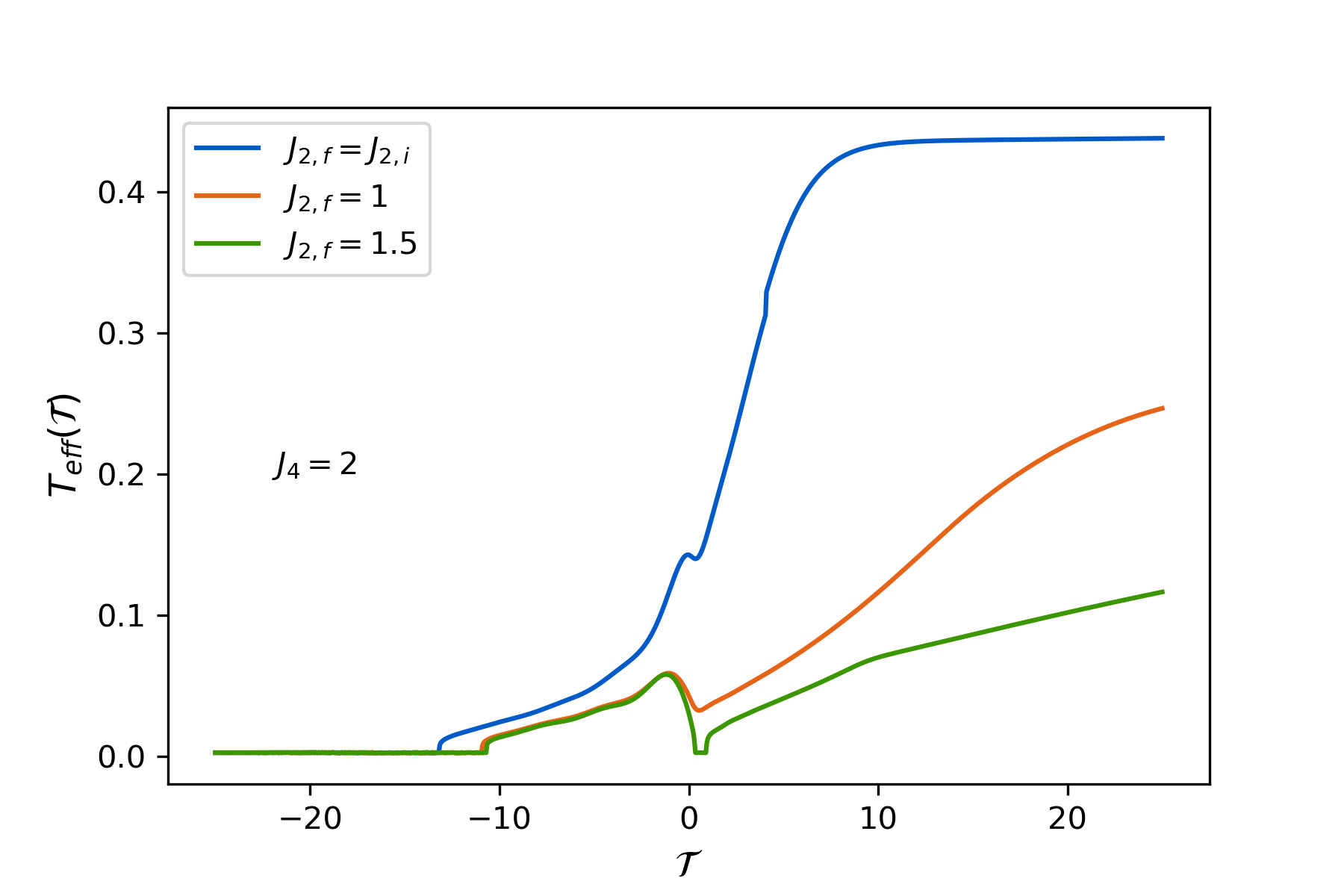}
	\caption{Dynamics of the effective temperature $T_{\rm eff}(\mathcal{T})$ after a double quench of $J_2$ and $J_4$. We consider a sudden switching-on of $J_4$ from zero to $J_4=2$ and different quench values of $J_{2,f}$. We see that the heating dynamics of the system slows down upon increasing the strength of the $J_2$ quench, a sign of extended prethermalization dynamics.}\label{fig7}
\end{figure}

\subsection{Dynamics of Effective Temperature and Real-time Decay of Retarded Green's function}

In order to discuss the effect of this double quench we focus on the dynamics of the effective temperature, as obtained from the low-frequency regime of the distribution function as discussed in Sec.~\ref{sec:results1}. In particular we plot in Fig.~\ref{fig7} the effective temperature for a quench to $J_4=2$ and different values of the $J_{2,f}$ parameter. We see that quenching the $J_2$ term leads to a significant slow down of the effective temperature dynamics as compared to the case in which $J_2$ is kept constant (top line in Fig.~\ref{fig7}), and already for $J_{2,f}=1.5$ the effective temperature does not reach a stationary state on the time scales of our simulation. In addition to a slowing down of the thermalization time we also note that the long-time limit of the effective temperature is reduced by the quench, namely the system heats up less. 

This can be understood by noticing that, at least for $J_{2,f}\gg J_{4}$ when the interactions can be neglected to a leading order, the final and initial Hamiltonian are both solvable SYK$_2$ models, a regime in which the dynamics is known to equilibriate instantaneously without heating production.

\begin{figure}[t]
	\includegraphics[width=0.45\textwidth]{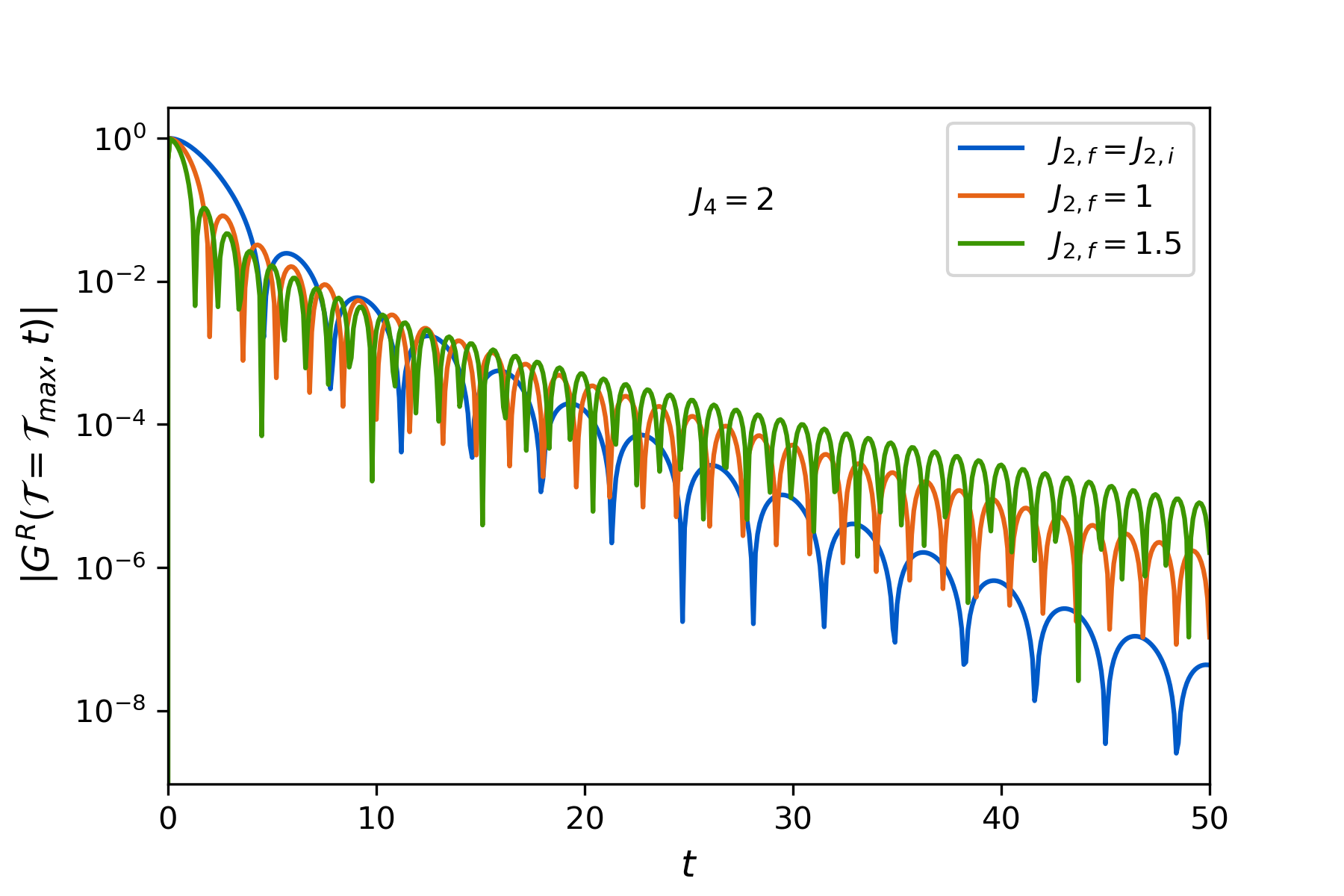}
	\caption{Decay of real-time retarded Green's function $G^R(\mathcal{T},t)$, long after the quench, as a function of the relative time $t$ after a double quench of $J_2$ and $J_4$. We consider a sudden switching-on of $J_4$ from zero to $J_4=2$ and different quench values of $J_{2,f}$. We see that the decay rate is slower upon increasing the quench of $J_2$.}\label{fig8}
\end{figure}

The slow-down of the dynamics in presence of a $J_2$ quench can be also clearly seen from the decay in time of the retarded Green's function long time after the quench, as discussed for a single quench in Sec.~\ref{sec:results1}.  In Fig.~\ref{fig8} we plot  $G^R(\mathcal{T},t)$ long time after the quench $\mathcal{T}=25$ as a function of the relative time $t$, after a quench to $J_4=2$ and different values of the $J_2$ quench. We see that the decay is always exponential, as in the case of a single quench discussed in Eq.~(\ref{eqn:Gammainf}) although with a rate $\Gamma_{\infty}$ (not shown), which decreases upon increasing $J_{2,f}/J_{2,i}$.

\section{Discussion: Quench-Induced NFL to FL Crossover}\label{sec:discussion}

We now summarize our results for the quench dynamics of the mixed SYK$_4$+SYK$_2$ model.  Our analysis so far has focused on the dependence of the nonequilibrium dynamics from the quench parameters, respectively the strength of the interaction $J_4$ and of the single particle quench $J_{2,f}/J_{2,i}$. An interesting question we would like to address here is how the long time limit of the quench problem relates to the equilibrium properties of the mixed SYK model and whether in particular it is possible to explore the crossover from NFL to FL in the quench dynamics.

\begin{figure}[t]
	\includegraphics[width=0.45\textwidth]{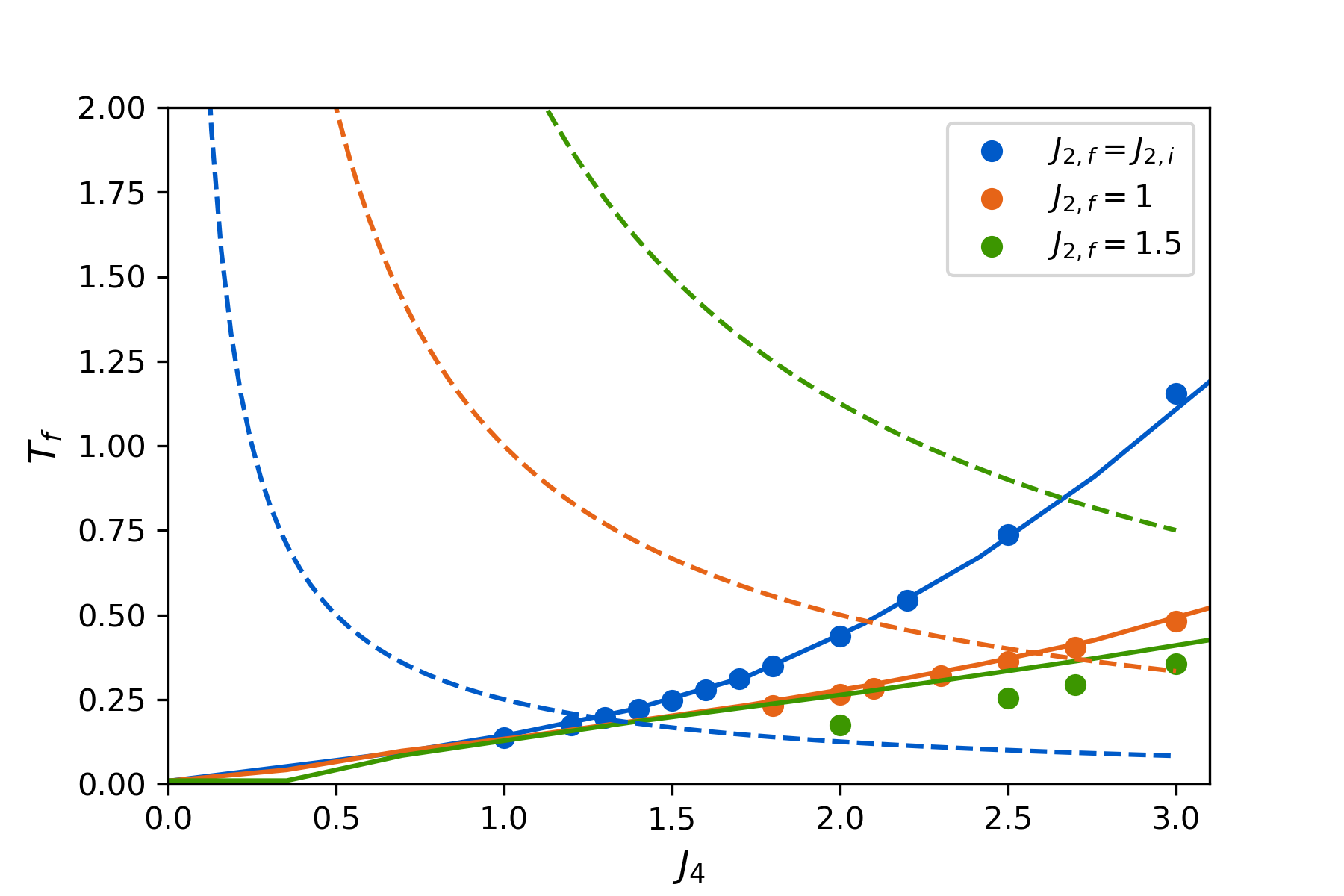}
	\caption{Quench phase diagram for the mixed SYK$_4$+SYK$_2$ model. Effective final temperature versus the quench parameter $J_4$ and for different quenches of the $J_2$ term. We see that quite generically the effective temperature production is slower upon quenching also $J_2$. For comparison we plot in the same plane the crossover scale $T^*\sim J_{2,i}^2/J_4$ and $T^*\sim J_{2,f}^2/J_4$ corresponding to the equilibrium FL-to-NFL crossover. }\label{fig9}
\end{figure}

To this extent it is useful to plot a dynamical phase diagram for the quench problem, as in Fig.~\ref{fig9}, where we show the dependence of the effective temperature $T_f$ from the quench parameter $J_4$ for different values of the $J_2$ quench, corresponding respectively to a single quench ($J_{2i}=J_{2f}$) and to a double quench  ($J_{2i}\neq J_{2f}$). The effective temperature is obtained from the dynamics of the distribution function (points in Fig.~\ref{fig9}), as discussed in Sec.~\ref{sec:results1}, and further compared to the estimate obtained from total energy conservation (lines in Fig.~\ref{fig9}, as discussed in App.~\ref{app:equilibrium}). We see that $T_{f}$ increases with $J_4$, namely the stronger is the value of the interaction quench the more the system heats up, however this increase is slower for a double quench protocol.  Specifically, we see that upon increasing the single particle coupling term $J_{2f}/J_{2i}>1$, at fixed $J_4/J_{2,i}$ ratio, the final temperature at which the system effectively thermalizes is lower as compared to a pure $J_4$ quench. This can be easily understood in the regime $J_4\ll J_{2,i}$, when essentially the dynamics reduces to a quench between SYK$_2$ models which are known to not produce substantial heating, but remains true even at intermediate values of $J_4$ as our numerical results show. This result indicates that the out of equilibrium physics of the mixed SYK model is much richer than the pure SYK$_4$ and that changing the quench protocol allows to explore different dynamical regimes.
This result is particularly interesting in connection with the scale $T^*\sim J_{2,f}^2/J_4$ that controls the equilibrium properties of the final mixed SYK model and in particular its NFL-to-FL crossover. For comparison, we plot these energy scales on the same plane as the effective final temperature, see Fig.~\ref{fig9}. For a single quench of the interaction term $J_4$ we see that most of our data lie above the crossover scale and therefore we expect to see a behavior compatible with a pure SYK$_4$. On the other hand, for a double quench the effective temperature decreases and the crossover line is also pushed to larger values of $J_4$. As a result, quenching the single particle scale we expect to see a behavior which is more compatible with a pure SYK$_2$. 

\begin{figure}[t]
	\includegraphics[width=0.45\textwidth]{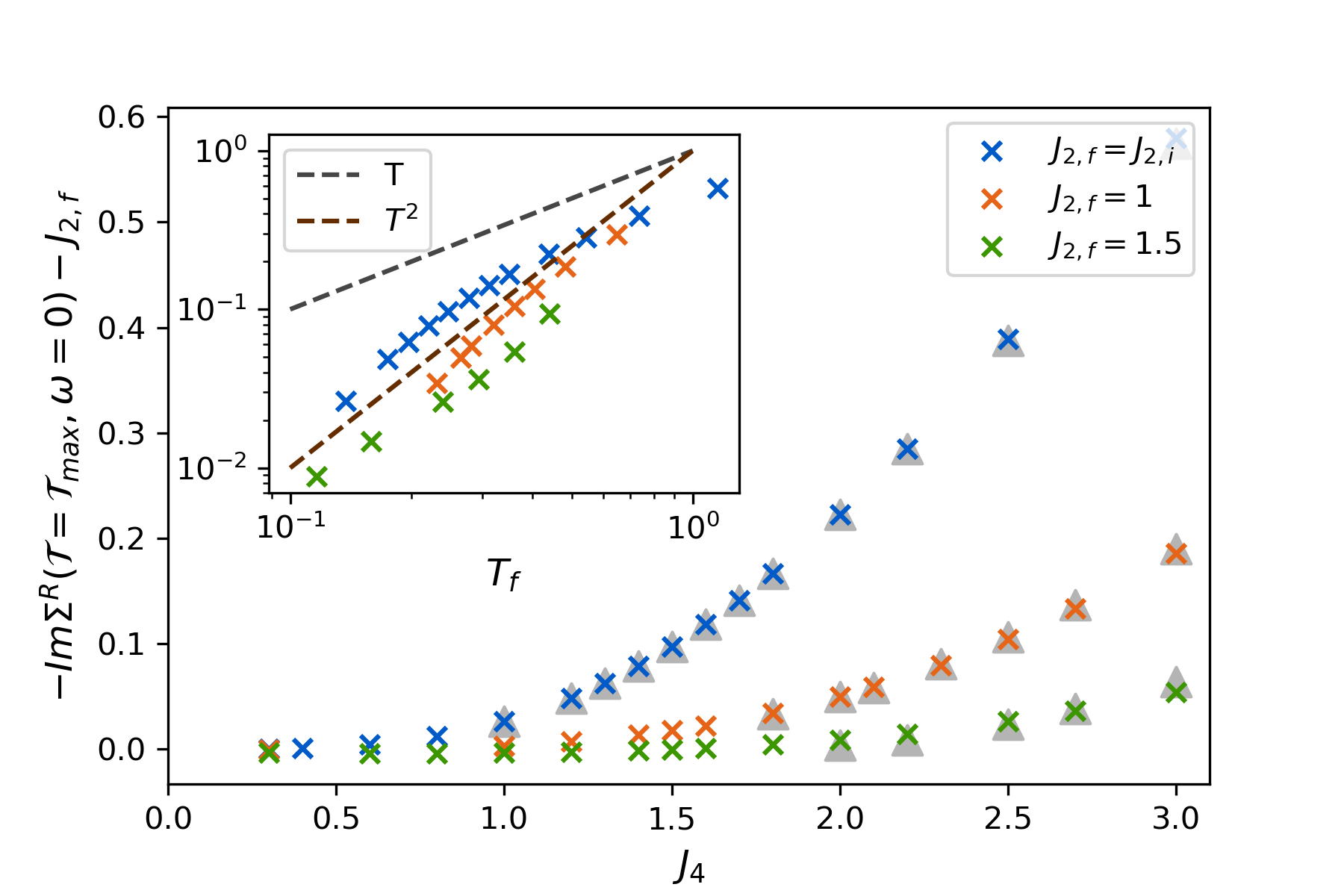}
	\caption{Out of equilibrium scattering rate, defined as the imaginary part of the retarded self-energy long time after the quench, $ - \textrm{Im} \, \Sigma^{R}(\mathcal{T} = \mathcal{T}_{\rm max}  ,\omega = 0) $, as a function $J_{4}$ for a single quench (blue points) or for a double quench to $J_{2,f}\neq J_{2,i}$ (orange and green points).  Grey triangles: equilibrium result at $T_{f}$. Inset: same quantity plotted versus the effective temperature $T_f$, showing a linear scaling and deviations from it at lower effective temperatures.	}\label{fig10}
\end{figure}
To confirm this expectation we  consider the out of equilibrium scattering rate for the Majorana fermions, namely the imaginary part of the retarded self-energy long time after the quench, $- \textrm{Im} \, \Sigma^{R}(\mathcal{T} = \mathcal{T}_{\rm max},\omega = 0) $.  This can be readily obtained from the nonequilibrium Green's functions through Eq.~\ref{eq:sigmaRA} and \ref{eq:sigmaalphabeta} and after Wigner transform. In thermal equilibrium this quantity is known to be sensitive to the crossover scale $T^*$, as we discuss in the Appendix~\ref{app:equilibrium}. In Fig.~\ref{fig10} we plot this quantity as a function of $J_4$ for different values of $J_{2,f}$ and $J_{2,i}$, including therefore both the single quench protocol discussed in Sec.~\ref{sec:results1} as well as the double quench. As expected the scattering rate, much like the effective temperature, increases with $J_4$ but with a rate that depends strongly on $J_{2,f}/J_{2,i}$ and decreases for large quenches of the $J_2$ coupling. To further confirm the overall thermalization of the nonequilibrium dynamics we compare this quantity with its equilibrium version at the final effective temperature (see grey triangles) finding perfect agreement. Furthermore in the inset of Fig.~\ref{fig10} we plot the same scattering rate as a function of the effective temperature $T_f$ (in log scale) for the three values of  $J_{2,f}/J_{2,i}$  considered. 
We see that for a double quench when the effective temperature decreases while the crossover scale is pushed towards higher values, we are able to see more clearly a behavior which is consistent with the FL scaling 
\begin{align}\label{eqn:Sigma_FL}
- \textrm{Im} \, \Sigma^{R}\sim T^2_f\,.
\end{align}

On the other hand for a single quench, corresponding to $J_{2,f}=J_{2,i}$, when the effective temperature is above the crossover scale (See Fig.~ \ref{fig9} ) the scattering rate deviates from the FL scaling of Eq.~(\ref{eqn:Sigma_FL}) and is compatible with the behavior expected from SYK$_4$ at intermediate temperatures (see Appendix~\ref{app:equilibrium}), when the square-root low temperature regime crosses over to a linear scaling.

We emphasize that reaching lower effective temperatures in our dynamical approach is challenging since the dynamics slow down significantly and the long time stationary limit becomes unaccessible to our finite time simulation. Nevertheless based on the evidence in Fig.~\ref{fig10}  we can safely conclude that the scattering rate is a good probe of the quench-induced crossover between NFL and FL. In conclusion we note that, on the other hand, the relaxation rate $\Gamma_{\infty}$ defined from the long-time limit of the retarded Green's function is not a good probe of the crossover already in equilibrium, as we show explicitly in Appendix ~\ref{app:equilibrium}.

An interesting point that we leave for future work is whether one can draw a formal connection between probes of thermalization and chaos obtained through the Liapunov exponent of out-of-time ordered correlators and the scattering rate obtained through single particle self-energies and Green's functions.

\section{Conclusions}\label{sec:Conclusions}

In this work we have discussed the quench dynamics of the mixed SYK$_4$+SYK$_2$ model in the large $N$ limit. Specifically we have solved the real-time KB equations numerically for two different quench protocols, corresponding to a sudden switching of the $J_4$ interaction and a simultaneous quench of the single-particle bandwidth $J_2$. 
We have compared the full KB dynamics to an approach based on a Quantum Boltzmann equation, derived starting from the exact large $N$ Dyson equation  without any quasiparticle approximation. We have shown that this approach, which relies on a gradient expansion for a slow average time is able to capture correctly the long-time dynamics of the problem.

Our results have shown that quite generically the unitary dynamics of this model thermalizes to a finite temperature thermal equilibrium state, as confirmed by both the spectral function and the distribution functions of the Majorana modes, two quantities that however evolve on much different time scales.  In particular the dynamics of the effective temperature as obtained from the effective FDT on the distribution function appears to become very slow for weak quenches, or equivalently for large quenches of the single particle term $J_2$. We have connected this result to the onset of prethermalization in the quench dynamics of this system. 

As compared to quenches in the pure SYK$_4$ model, the mixed case enjoys a much richer dependence from the quench parameters, encapsulated in the nonequilibrium phase diagram shown in Fig.~\ref{fig9}. We have shown that quite generically a quench of the $J_4$ coupling leads to a finite temperature which can be above or below the crossover scale $T^*$ and that quenching the single particle term $J_2$ significantly reduce the heating in the system and allows to access the NFL to FL crossover through the quench dynamics, as we have shown by looking at the nonequilibrium scattering rate at long times.

Our results offer therefore a complementary picture, based on the full out of equilibrium dynamics, to the studies on the scrambling properties of the mixed SYK model and point out a possible interesting connection between slow scrambling and prethermalization that could be worth discussing further in the future.  Further perspectives opened by this work includes the investigation of different nonequilibrium settings involving mixed SYK-like models, such as those recently considered in connection with traversable wormholes in the high-energy literature. Finally we note that there is currently large interest in disordered fully connected models with mixed competing interactions which, even at the classical level, have been shown to possess intriguing properties~\cite{folena20202rethinking}.

\begin{acknowledgements}
We thank A. Georges  for helpful discussions and the Coll\`ege de France IPH cluster for computational resources.
This work was supported by the ANR grant ”NonEQuMat” (ANR-19-CE47-0001).
\end{acknowledgements}

\appendix
\section{Equilibrium Properties of the Mixed SYK Model}
\label{app:equilibrium}

\begin{figure}[t]
	\includegraphics[width=0.45\textwidth]{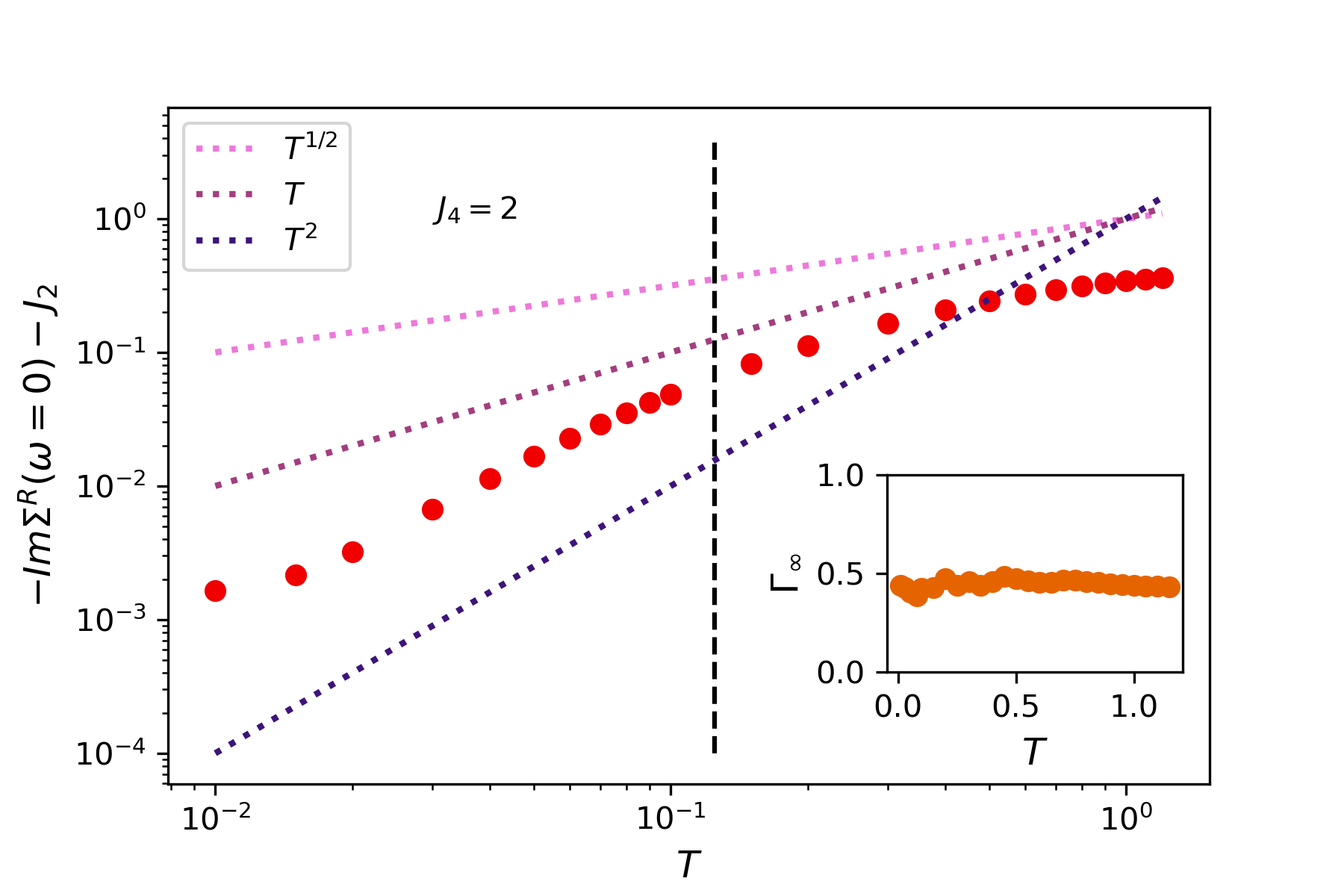}
	\caption{Equilibrium scattering rate, defined as the imaginary part of the retarded self-energy in equilibrium at zero frequency, $ - \textrm{Im} \, \Sigma^{R}(\omega = 0) $, as a function of temperature. We see the crossover from FL ($ - \textrm{Im} \, \Sigma^{R}\sim T^2$) to NFL ($ - \textrm{Im} \, \Sigma^{R}\sim\sqrt{T}$) scaling, as the temperature is raised above $T^*$ (dashed line), with an almost linear regime around the crossover. For comparison we plot in the inset the decay rate of the equilibrium retarded Green's function, equivalent to $\Gamma_{\infty}$ defined in Sec.~\ref{sec:results1}, that show a featureless dependence from the temperature. }\label{fig11}
\end{figure}

In this section we briefly recall some of the equilibrium properties of the mixed SYK$_4$+SYK$_2$ model in the large $N$ limit. In this case the Dyson equation for the retarded single particle Green's function can be written directly in frequency as
\begin{equation}
    G^R(\omega)=\frac{1}{\omega-\Sigma^R(\omega)}
\end{equation}
where the retarded self-energy can be still written in the time-domain as 
\begin{equation}\label{eq:sigmaR_eq}
    \Sigma^R(t)=-\frac{J^2_4}{4}\left(G^R(t)^3+3G^R(t)G^K(t)^2\right)+J^2_2G^R(t)\,
\end{equation}
where the Keldysh component $G^K(t)$ is related to the retarded one by the fluctuation-dissipation theorem, Eq.~\ref{eq:fdt} of the main text. The two equations above can be solved iteratively, going back and forth from the frequency to the time domain, until a converged solution is found. A key feature of the equilibrium SYK$_4$+SYK$_2$ model is the crossover from Non-Fermi-Liquid to Fermi-Liquid scaling as temperature of the system is lowered below the scale $T^*\sim J_2^2/J_4$. This crossover can be clearly seen in the equilibrium scattering rate of the Majorana fermions, given by the imaginary part of the retarded self-energy at zero frequency, that we plot in Fig.~\ref{fig11} as a function of temperature. We see that the low-temperature $~T^2$ behavior crosses over to a square root scaling $\sim \sqrt{T}$, expected from SYK$_4$,  when the temperature is above the dashed line, indicating the crossover scale $T^*$ at that values of $J_2,J_4$, with an apparent linear regime around $T^*$.
 At higher temperatures instead the scattering rate saturates. We further note that in Fig.~\ref{fig11} the scattering rate is shifted with respect to a constant value proportional to $J_2$, the single particle bandwidth, that is also responsible for a finite imaginary part of the self-energy from Eq.~\ref{eq:sigmaR_eq}, although not related to the many body interactions. For comparison we plot in the inset another measure of the decay rate, obtained from the long-time decay of the retarded Green's function, analogous to $\Gamma_{\infty}$ defined in Sec.~\ref{sec:results1}. This quantity on the other hand has a very weak dependence on temperature and does not show any signature of the crossover scale.

\subsection{Effective temperature from energy conservation}

We now discuss how to estimate the final effective temperature to which the system thermalizes after a given quench protocol, from the knowledge of the equilibrium thermodynamics of the mixed SYK model. The total energy after the quench is in fact conserved and equal to 
\begin{equation}
    E_{Q} = \langle H_{f} \rangle_{H_{i},T_i}
\end{equation}
where $\langle\,\rangle_{H_{i},T_i}$ indicates thermal averages over the initial Hamiltonian $H_i$ at temperature $T_i$ and $H_f$ is the Hamiltonian driving the evolution after the quench. On the other hand, if the system thermalizes one could expect the energy density to be given by the thermal expectation value of the final Hamiltonian at a temperature $T_f$, i.e.
\begin{equation}\label{eqn:Teff_energycons}
    E_{Q} = E_{T}(T_{f})= \langle H_{f} \rangle_{H_{f}, T_{f}}
\end{equation}
where $\langle\,\rangle_{H_{f},T_f}$ indicates thermal averages over the final Hamiltonian $H_f$ at temperature $T_f$. Eq.~(\ref{eqn:Teff_energycons}) allows to estimate the effective temperature from the knowledge of the energetics of the problem. The averages entering the equations above can be found by using functional derivatives and read
\begin{equation}
    E_{Q} = i \frac{J_{2,i} J_{2,f}}{2} \int_{0}^{\infty} d t \big( G_{i}^{>}(t)^{2} - G_{i}^{<}(t)^{2} \big)
\end{equation}
\begin{align}
    E_{T}(T_{f})& = i \frac{J_{2,f}^{2}}{2} \int_{0}^{\infty} d t \big( G_{f}^{>}(t)^{2} - G_{f}^{<}(t)^{2}  \big) +\\\nonumber
    &- i \frac{J_{4}^{2}}{2} \int_{0}^{\infty} d t \big( G_{f}^{>}(t)^{4} - G_{f}^{<}(t)^{4}  \big)
\end{align}
where $ G_{i} = G_{\textrm{SYK}_{2}, \, T=0} $ and $G_{f} = G_{\textrm{SYK}_{4} + \textrm{SYK}_{2},\, T_{f}}$. In Figure \ref{fig3} the solid line is the curve $T_{f} = T_{f}(J_{4})$ obtained by this method.

\section{Quantum Boltzmann Equation}\label{app:QBE}

In this section we sketch the derivation of the Quantum Boltzmann equation for the mixed SYK model. We start from the real-time Dyson equation~(\ref{eqn:Dyson}) in the main text, that we write for the Keldysh component of the Green's function as
\begin{equation}\label{eqn:dysonK}
   \Big( [G_{0}^{R}]^{-1} - \Sigma^{R} \Big) \circ G^{K} = \Sigma^{K} \circ G^{A}
   \end{equation}
where $\circ$ is the real-time convolution and  $  [G_{0}^{R}]^{-1} = i \delta(t-t^{\prime}) \partial_{t} $ the non-interacting Majorana retarded Green's function. We then parametrize the Keldysh Green's function in terms of the distribution function $F(t_{1},t_{2})$ as $G^{K} = G^{R} \circ F - F \circ G^{A}$ and recast Eq.~(\ref{eqn:dysonK}) into a quantum kinetic equation form \cite{kamenev_2011}
\begin{equation}
    [\partial_{t_{1}},F ] = i \Sigma^{K} - i \big( \Sigma^{R} \circ F - F \circ \Sigma^{A} \big)
\end{equation}
where $ [ \partial_{t_{1}}, F ](t_{1},t_{2}) = (\partial_{t_{1}} + \partial_{t_{2}}) F(t_{1},t_{2}) $. To take the Wigner transform of this equation we need to take care of the convolution. The Wigner transform of $f = g \circ h$ is \cite{kamenev_2011}
\begin{align}\label{eq:grad_exp}
    f(\mathcal{T},\omega) &= g(\mathcal{T},\omega) e^{- \frac{i}{2} (\partial_{\mathcal{T}}^{\leftarrow} \partial_{\omega}^{\rightarrow} - \partial_{\omega}^{\leftarrow} \partial_{\mathcal{T}}^{\rightarrow})} h(\mathcal{T},\omega) \\
    &= g(\mathcal{T},\omega) h(\mathcal{T},\omega) - \frac{i}{2} \big[ \partial_{\mathcal{T}} g \partial_{\omega} h - \partial_{\omega} g \partial_{\mathcal{T}} h \big] + \cdots
\end{align}
We made a gradient expansion in the second line, valid for large central time $\mathcal{T}$. To lowest order in the gradient expansion, the kinetic equation reads
\begin{equation}\label{eqn:gradient_zero}
    \partial_{\mathcal{T}} F(\mathcal{T},\omega) = i \Sigma^{K}(\mathcal{T},\omega) - i F(\mathcal{T},\omega) \big( \Sigma^{R}(\mathcal{T},\omega) - \Sigma^{A}(\mathcal{T},\omega) \big)
\end{equation}
After the quench the self-energies $ \Sigma^{K}(t_{1},t_{2}),  \Sigma^{R}(t_{1},t_{2}) - \Sigma^{A}(t_{1},t_{2}) $ are (where we omit the time-dependence for simplicity)
\begin{align}\label{eqn:SigmaRAK}
    &\Sigma^{K} = J_{2,f}^{2} G^{K} - \frac{J_{4}^{2}}{4} \big[ \left(G^{K}\right)^{3} + 3 G^{K} \big( G^{R} - G^{A} \big)^{2} \big]\\
 &\Sigma^{R} - \Sigma^{A} = J_{2,f}^{2} \big( G^{R} - G^{A} \big) - \frac{J_{4}^{2}}{4} \big[ \big( G^{R} - G^{A} \big)^{3} +\nonumber\\
&+ 3 \big( G^{R} - G^{A} \big) \left(G^{K}\right)^2 \big]
\end{align}
Taking the Wigner transform of these expressions and using the fact that  $ i G^{K}(\mathcal{T},\omega) = A(\mathcal{T},\omega) F(\mathcal{T},\omega) $ and $ G^{R}(\mathcal{T},\omega) - G^{A}(\mathcal{T},\omega) = 2 i \textrm{Im} G^{R}(\mathcal{T},\omega) = - i A(\mathcal{T},\omega) $ we obtain for the two self-energies $ \Sigma^{K}, \Sigma^{R}-\Sigma^A$ entering Eq.~(\ref{eqn:gradient_zero}) the results
\begin{widetext}
\begin{equation}
\begin{split}\label{eqn:SigmaRAK_wign}
    \Sigma^{K}(\mathcal{T},\omega) &= J_{2,f}^{2} G^{K}(\mathcal{T},\omega) - i \frac{J_{4}^{2}}{4} \int \frac{d \omega_{1}}{2 \pi} \frac{d \omega_{2}}{2 \pi} A(\mathcal{T}, \omega - \omega_{1} - \omega_{2}) A(\mathcal{T},\omega_{1}) A(\mathcal{T},\omega_{2}) \\
    &\times \big[ F(\mathcal{T},\omega - \omega_{1} - \omega_{2}) F(\mathcal{T},\omega_{1}) F(\mathcal{T},\omega_{2}) + 3 F(\mathcal{T},\omega - \omega_{1} - \omega_{2}) \big]\\
\Sigma^{R}(\mathcal{T},\omega) - \Sigma^{A}(\mathcal{T},\omega) &= - i J_{2,f}^{2} A(\mathcal{T},\omega) - i \frac{J_{4}^{2}}{4} \int \frac{d \omega_{1}}{2 \pi} \frac{d \omega_{2}}{2 \pi} A(\mathcal{T}, \omega - \omega_{1} - \omega_{2}) A(\mathcal{T},\omega_{1}) A(\mathcal{T},\omega_{2}) \\
    &\times \big[ 1 + 3 F(\mathcal{T},\omega_{1}) F(\mathcal{T},\omega_{2})  \big]
\end{split}
\end{equation}
\end{widetext}
%
Plugging these into Eq.~(\ref{eqn:gradient_zero}) we finally obtain a quantum kinetic equation for the distribution function, i.e.
\begin{equation}
    \partial_{\mathcal{T}} F(\mathcal{T},\omega) = I^{\textrm{coll}} \big[ F(\mathcal{T},\omega) \big]
\end{equation}
where the collision integral $ I^{\textrm{coll}} $ can be re-written after some simple manipulations as
\begin{widetext}
\begin{align}\label{eqn:Icoll_explicit}
     I^{\textrm{coll}} \big[ F(\mathcal{T},\omega) \big] = \frac{J_{4}^{2}}{4} \int \frac{d \omega_{1}}{2 \pi} \frac{d \omega_{2}}{2 \pi} &A(\mathcal{T},\omega - \omega_{1} - \omega_{2}) A(\mathcal{T},\omega_{1}) A(\mathcal{T},\omega_{2}) \times \big[ F(\mathcal{T},\omega - \omega_{1} - \omega_{2}) F(\mathcal{T},\omega_{1}) F(\mathcal{T},\omega_{2})+\nonumber \\
    &+ F(\mathcal{T},\omega - \omega_{1} - \omega_{2}) + F(\mathcal{T},\omega_{1}) + F(\mathcal{T},\omega_{2}) - F(\mathcal{T},\omega) \big( 1 + F(\mathcal{T},\omega_{1}) F(\mathcal{T},\omega_{2})+\nonumber \\
    &+ F(\mathcal{T},\omega - \omega_{1} - \omega_{2}) F(\mathcal{T},\omega_{1}) + F(\mathcal{T},\omega - \omega_{1} - \omega_{2}) F(\mathcal{T},\omega_{2}) \big) \big]
\end{align}
\end{widetext}
Interestingly we note that the term in the self-energies proportional to $J_{2,f}^{2}$ in Eq.~(\ref{eqn:SigmaRAK_wign}) drops out of the expression for the collision integral $I^{\textrm{coll}}[F]$, at least to the lowest order. This underlies the fact that the $J_2$ coupling, while resulting in a self-energy contribution for the Majorana mode after disorder averaging, does not provide an effective scattering mechanism. This is a further  indication that the pure SYK$_2$ is not able to reach thermal equilibrium by itself.  We also note that the collision integral in Eq.~(\ref{eqn:Icoll_explicit}) has a specific dependence from the distribution function $F(\mathcal{T},\omega)$. In fact one can show, using the properties of the hyperbolic tangent function, that  for an equilibrium distribution function the collision integral  vanishes exactly, i.e.
\begin{equation}
    I^{\textrm{coll}} \big[ F^{eq}(\omega) \big] = 0 \quad \textrm{where} \quad F^{eq}(\omega) = th \Big( \frac{\beta \omega}{2} \Big)
\end{equation}
a properties which guarantees that the QBE dynamics leads to thermal equilibrium. We note however that the temperature $\beta$ in the above equation is only fixed by the initial condition. 
As for the distribution function, we can also write an equation for the spectral density $A(\mathcal{T},\omega)$. We start from the left and right Dyson equations for the retarded component of the Green's function
\begin{align}
    G_{0}^{-1} \circ G^{R} &= 1 + \Sigma^{R} \circ G^{R} \\
    G^{R} \circ G_{0}^{-1} &= 1 + G^{R} \circ \Sigma^{R}
\end{align}
We substract the two equations:
\begin{equation}
    \big[G_{0}^{-1} - \Sigma^{R}, G^{R} \big] = 0
\end{equation}
Likewise for the advanced component:
\begin{equation}
    \big[G_{0}^{-1} - \Sigma^{A}, G^{A} \big] = 0
\end{equation}
We can rewrite these two equations in terms of $ G_{\pm } = ( G^{R} \pm G^{A}  ) / 2  $ and $ \Sigma_{\pm } = ( \Sigma^{R} \pm \Sigma^{A}  ) / 2  $:
\begin{align}
    \big[ G_{0}^{-1}- ( \Sigma_{+} + \Sigma_{-}), G_{+} + G_{-}   \big] &= 0\\
    \big[ G_{0}^{-1}- ( \Sigma_{+} - \Sigma_{-}), G_{+} - G_{-}   \big] &= 0
\end{align}
The difference of these two equations is
 \begin{equation}
     \big[ G_{0}^{-1}, G_{-} \big] = \big[ \Sigma_{+} , G_{-} \big] + \big[ \Sigma_{-} , G_{+} \big]
 \end{equation}
We now take the Wigner transform of this equation and keep only the lowest order terms in the gradient expansion. According to equation (\ref{eq:grad_exp}) the Wigner transform of a commutator (for the convolution) is zero at lowest order, so the right-hand side yields zero. As $ G_{-}(\mathcal{T}, \omega) = - i A(\mathcal{T},\omega) / 2 $, we have
\begin{equation}
    \partial_{\mathcal{T}} A(\mathcal{T},\omega) = 0
\end{equation}
This is consistent with the results obtained by solving numerically the Kadanoff-Baym equation. The gradient expansion is valid only for large central time $ \mathcal{T}$ and we have in the numerics that the spectral function $A(\mathcal{T},\omega)$ equilibriates shortly after the quench while the distribution function $F(\mathcal{T},\omega)$ takes longer time.\\

\subsection{Approximate Solution of QBE for the mixed SYK model}

In the main text we have presented the full QBE numerical solution starting from an initial condition obtained through a short-time propagation of KB equations. This hybrid approach, which we have shown to compare well with the exact KB dynamics for the mixed SYK model, could be particularly useful to study systems with slow dynamics, which can be challenging to tackle with a full KB simulation on very long time scales. Still, the computational effort associated with a QBE simulation remains substantial due to the frequency convolution structure of the collision integral in Eq.~(\ref{eqn:Icoll_explicit}). A possible approach, developed in Ref.~\cite{tavora2013quench} to study quenches of interacting bosonic field theories, is to focus on the low-frequency part of the distribution function and to parametrize it in terms of few time-dependent coefficients within a small $\omega$ expansion
\begin{equation}
    F(\mathcal{T},\omega) = \frac{\beta( \mathcal{T})}{2} \omega + C(\mathcal{T}) \omega^{3} + \cdots 
\end{equation}
leading to a set of simple rate equations for the effective temperature and other time-dependent parameters, whose numerical integration could be performed very efficiently. In Ref.~\cite{tavora2013quench} this approach was shown to capture qualitatively well the QBE dynamics. In the case of the mixed SYK model however we found this simple approximation to break down. This is due to the fact that a low-frequency expansion for the distribution function only holds if
%
\begin{equation}
    \omega \ll \Delta_{T} \sim \textrm{min}(T_{i},T_{f})
\end{equation}
At the same time the structure of the collision integral is such that the distribution function  always appears multiplied by the spectral function $A$ at the same frequency. 
\begin{figure}[t!]
    \centering
    \includegraphics[scale = 0.5]{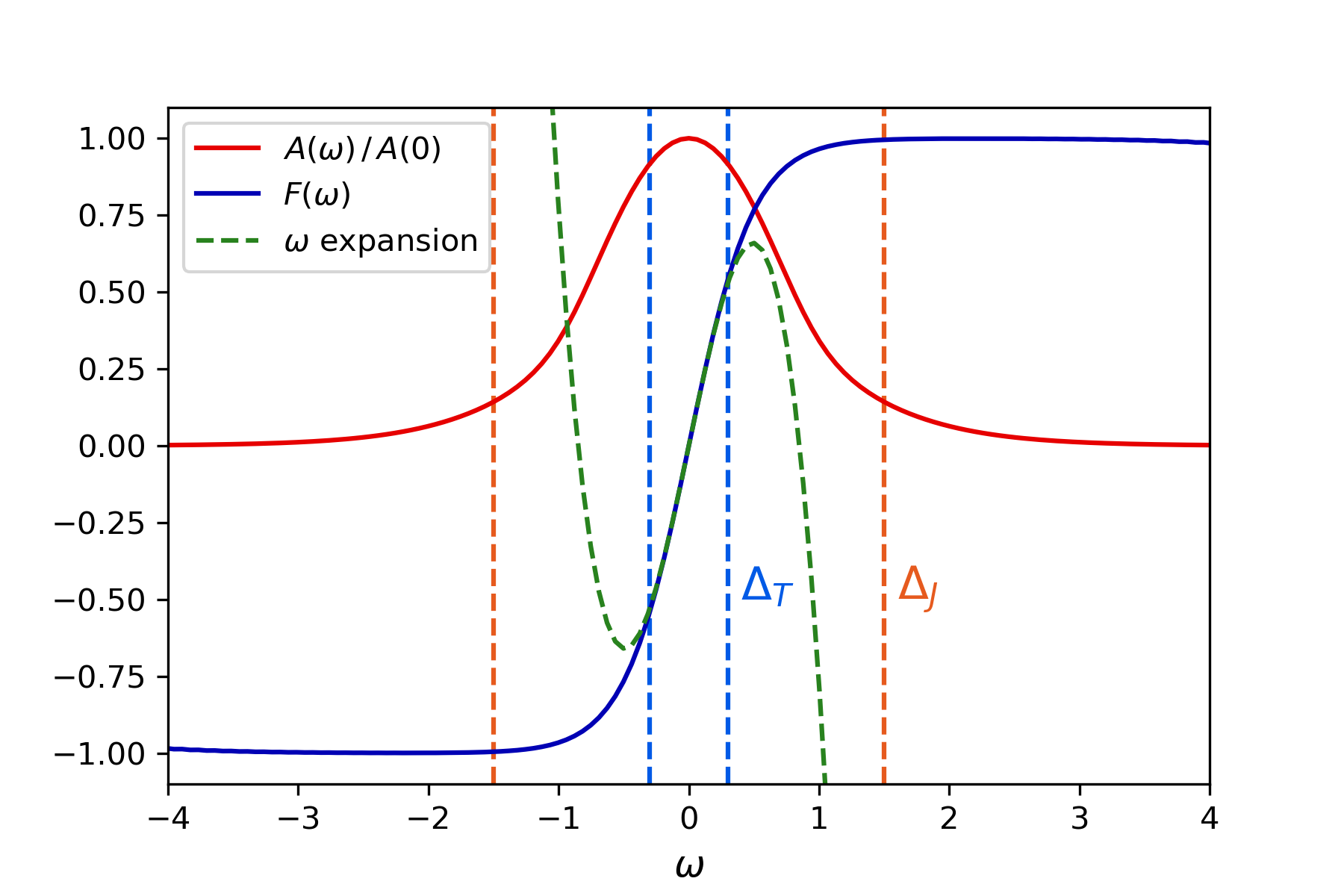}
    \caption{Frequency dependence of spectral and distribution functions at long times  $ \mathcal{T} = \mathcal{T}_{max} $ for the mixed SYK model after a quench to   $ J_{2} = 0.5 $,  $J_{4} = 1.5$, leading to a final temperature  $T_{f} \simeq 0.25$.}
    \label{figomegaexp}
\end{figure}
Thus, it is a good approximation to expand $F$ in powers of $\omega$ if the bandwidth $\Delta_{J}$ of $A$ is smaller or of the order of $\Delta_{T}$. This is typically the case when sharp quasiparticle excitations dominates the spectrum. On the other hand for SYK models we have typically very broad spectral functions with bandwidth  $\Delta_{J} \sim J  \gg \Delta_{T} $. This is shown for concreteness  in Fig.~(\ref{figomegaexp}) where we compare spectral and distribution functions long time after a quench to $ J_{2} = 0.5 $,  $J_{4} = 1.5$, leading to a final temperature  $T_{f} \simeq 0.25$. This further confirms that quasiparticle like approximations to the QBE are likely not able to capture the dynamics of SYK models and one needs to resort to the full solution of the QBE.


%

\end{document}